\documentclass[a4paper,fleqn,usenatbib]{mnras}

\usepackage{newtxtext,newtxmath}

\usepackage[T1]{fontenc}
\usepackage{ae,aecompl}

\usepackage{graphicx}	
\usepackage{amsmath}	
\usepackage{amssymb}	
\usepackage{longtable}
\usepackage{pdflscape}	
\usepackage{array}

\usepackage[normalem]{ulem}

\title[BSNIP Spectroscopy of 247 SNe~Ia]{Berkeley Supernova Ia Program: Data Release of 637 Spectra from 247 Type Ia Supernovae}

\author[B. E. Stahl et al.]
{Benjamin E. Stahl,$^{1,2}$\thanks{E-mail: benjamin\_stahl@berkeley.edu}\thanks{Marc J. Staley Graduate Fellow}
WeiKang Zheng,$^{1}$
Thomas de Jaeger,$^{1}$\thanks{Bengier Postdoctoral Fellow}
Thomas G. Brink,$^{1}$
\newauthor
Alexei V. Filippenko,$^{1,3}$
Jeffrey M. Silverman,$^{4}$
S. Bradley Cenko,$^{5,6}$
Kelsey I. Clubb,$^{1}$
\newauthor
Melissa L. Graham,$^{7}$
Goni Halevi,$^{1,8}$
Patrick L. Kelly,$^{9}$
Io Kleiser,$^{10}$
Isaac Shivvers,$^{1}$
\newauthor
Heechan Yuk,$^{11}$
Bethany E. Cobb,$^{12}$
Ori D. Fox,$^{13}$
Michael T. Kandrashoff,$^{1}$
\newauthor
Jason J. Kong,$^{1}$
Jon C. Mauerhan,$^{14}$
Xianggao Wang,$^{15}$
and Xiaofeng Wang$^{16}$
\\
$^{1}$Department of Astronomy, University of California, Berkeley, CA 94720-3411, USA\\
$^{2}$Department of Physics, University of California, Berkeley, CA 94720-7300, USA\\
$^{3}$Miller Senior Fellow, Miller Institute for Basic Research in Science, University of California, Berkeley, CA 94720, USA\\
$^{4}$Samba TV, San Francisco, CA 94107, USA\\
$^{5}$Astrophysics Science Division, NASA Goddard Space Flight Center, MC 661, Greenbelt, MD 20771, USA\\
$^{6}$Joint Space-Science Institute, University of Maryland, College Park, MD 20742, USA\\
$^{7}$Department of Astronomy, University of Washington, Box 351580, Seattle, WA 98195, USA\\
$^{8}$ Department of Astrophysical Sciences, Princeton University, 4 Ivy Lane, Princeton, NJ 08544, USA\\
$^{9}$School of Physics and Astronomy, University of Minnesota, 116 Church Street SE, Minneapolis, MN 55455, USA\\
$^{10}$NASA Jet Propulsion Laboratory, 4800 Oak Grove Dr, Pasadena, CA 91109, USA\\
$^{11}$Department of Physics and Astronomy, University of Oklahoma, 440 W. Brooks St., Norman, OK 73019, USA\\
$^{12}$Department of Physics, The George Washington University, Washington, DC 20052, USA\\
$^{13}$Space Telescope Science Institute, 3700 San Martin Drive, Baltimore, MD 21218, USA\\
$^{14}$The Aerospace Corporation, 2310 E. El Segundo Blvd., El Segundo, CA 90245, USA\\
$^{15}$Department of Physics, Guangxi University, Nanning 530004, China\\
$^{16}$Physics Department and Tsinghua Center for Astrophysics, Tsinghua University, Beijing, 100084, China
}
\date{Accepted XXX. Received YYY; in original form ZZZ}

\pubyear{2019}

\begin{document}

\label{firstpage}
\pagerange{\pageref{firstpage}--\pageref{lastpage}}
\maketitle

\begin{abstract}
We present 637 low-redshift optical spectra collected by the Berkeley Supernova Ia Program (BSNIP) between 2009 and 2018, almost entirely with the Kast double spectrograph on the Shane 3~m telescope at Lick Observatory. We describe our automated spectral classification scheme and arrive at a final set of 626 spectra (of 242 objects) that are unambiguously classified as belonging to Type Ia supernovae (SNe~Ia). Of these, 70 spectra of 30 objects are classified as spectroscopically peculiar (i.e., not matching the spectral signatures of ``normal'' SNe~Ia) and 79 SNe~Ia (covered by 328 spectra) have complementary photometric coverage. The median SN in our final set has one epoch of spectroscopy, has a redshift of 0.0208 (with a low of 0.0007 and high of 0.1921), and is first observed spectroscopically 1.1 days after maximum light. The constituent spectra are of high quality, with a median signal-to-noise ratio of 31.8 pixel$^{-1}$, and have broad wavelength coverage, with $\sim 95\%$ covering at least 3700--9800~\AA. We analyze our dataset, focusing on quantitative measurements (e.g., velocities, pseudo-equivalent widths) of the evolution of prominent spectral features in the available early-time and late-time spectra. The data are available to the community, and we encourage future studies to incorporate our spectra in their analyses.
\end{abstract}




\section{Introduction}
\label{sec:introduction}

Supernovae (SNe) have proven themselves to be powerful probes of the dynamic nature of the Universe on scales ranging from stellar to cosmological. The class of objects known as Type Ia supernovae (SNe~Ia), which result from the thermonuclear explosions of carbon-oxygen white dwarfs in binary systems \citep[e.g.,][]{Hoyle1960,Colgate1969,Nomoto1984}, have been of particular interest to astrophysicists for many years.

Despite intensive study, many important details of SNe~Ia remain poorly understood, if at all \citep[for a review, see][]{Howell2011}. How do differences in initial conditions lead to the variation in properties observed among SNe~Ia? What are the physical details of the explosion mechanism(s)? Are the progenitor systems ``single-degenerate'' \citep{WhelanSD} or ``double-degenerate'' \citep{Webbink1984,IbenDD}, and how do they contribute to the observed variance in SN~Ia attributes? To answer these and other questions, numerous observations of SNe~Ia will undoubtedly be required --- preferably obtained and reduced in a thorough and consistent manner.

Despite these outstanding questions regarding SNe~Ia as astrophysical objects, they are highly prized for their large and \emph{relatively} homogeneous optical spectra and luminosities at peak brightness, though some differences do exist \citep[e.g.,][and references therein]{Filippenko1997}. To the extent that their peak luminosities are ``standardisable,'' SNe~Ia are excellent cosmological distance indicators. Accordingly, much effort has been expended in developing methods to better calibrate relationships between various observables and peak luminosity. The ``Phillips relation'' identifies a correlation between luminosity at peak brightness and light-curve decline rate for most SNe~Ia \citep{Phillips1993}. By making use of optical colours, \citet{Riess1996} have devised a method that yields further improvements, including the determination of the extinction caused by dust in the host galaxy of a SN~Ia. Distance measurements derived using such methods led to the discovery of the accelerating expansion of the Universe \citep{Riess1998,Perlmutter1999}, which revolutionised the field of cosmology. Indeed, the nature of the dark energy that gives rise to the acceleration is currently one of the most important questions in physics.

SNe~Ia have since been used to place increasingly stringent constraints on cosmological parameters \citep{Astier2006,Riess2007,Hicken2009,Suzuki2012,Betoule2014,Jones2018,Scolnic2018} and continue to provide precise measurements of the Hubble constant \citep{Riess2016,Riess2019,Dhawan2018}. As spectra must contain more information than light curves, many have searched for and identified spectroscopic parameters to make SN~Ia distance measurements more precise \citep{Bailey2009,Wang2009,Blondin2011,bsnipIII,Fakhouri2015,Zheng18}. In addition, \citet{Foley2011} found that the intrinsic colour of SNe~Ia at peak brightness depends on the velocity of their ejecta, and \citet{Wang13} have shown that the latter has a significant connection to SN~Ia birthplace environments --- and hence progenitor stars. It is likely that future increases in distance measurement precision will make use of spectroscopic parameters, motivating the need for extensive, consistent samples of SN~Ia spectra.

The Berkeley Supernova Ia Program (BSNIP) is a large-scale effort to study the properties of SNe~Ia at low redshift ($z \lesssim 0.05$), primarily via optical spectroscopy \citep[henceforth S12a]{bsnipI} and photometry \citep[][henceforth G10 and S19, respectively]{Ganeshalingam2010,S19a}. The spectra presented in this data-release paper are complementary to those published by S12a, and extend the BSNIP SN~Ia spectral dataset to cover the period from 1989 through 2018. Our strategy is generally to observe as many SNe~Ia as possible, with particular effort invested in obtaining frequent spectral coverage of peculiar objects. Furthermore, we strive for spectral coverage of all objects that our group is also observing photometrically (consequently, there is considerable overlap in SNe~Ia between the spectra presented herein and the photometric dataset released by S19), and we aim to provide prompt spectroscopic classifications of all SNe discovered by the 0.76-m Katzman Automatic Imaging Telescope at Lick Observatory \citep[KAIT;][]{kait}. Our spectra are obtained and reduced in a controlled and consistent manner, thereby eliminating many of the systematic differences that manifest when distinct datasets are collected into one sample.

In this data release, we present and characterise 637 optical spectra of 247 distinct objects collected by the BSNIP between the beginning of 2009 and the end of 2018. The spectra were obtained with the Shane 3~m telescope at Lick Observatory and the Keck-I 10~m telescope at the W. M. Keck Observatory. Of the full set of spectra, 546 are published here for the first time. When we combine our spectral dataset with that presented by S12a, we obtain a sample of nearly 2000 spectra of low-redshift SNe~Ia, all of which have been observed and reduced in a consistent manner. We organise the remainder of this paper as follows. Section~\ref{sec:data} describes the organisation, observation, and reduction strategies employed in assembling our dataset. In Section~\ref{sec:classification} we detail our spectral classification scheme, and we study its results and derive final object classifications. We present our final spectroscopic dataset and explore its early-time and late-time evolution in Section~\ref{sec:results}, and we conclude with Section~\ref{sec:conclusion}.

\section{Data}
\label{sec:data}

\subsection{Data Management and Selection}
\label{ssec:selection}

All BSNIP spectroscopy, along with useful metadata for those observations and the SNe in them (e.g., observer, reducer, host galaxy, redshift, etc.), are catalogued in our UC Berkeley SuperNova DataBase\footnote{\url{http://heracles.astro.berkeley.edu/sndb/}} \citep[SNDB; S12a,][]{Shivvers16} after the data are processed and reduced (see Section~\ref{sec:data_reduction} for a summary of our data-processing techniques). Therefore, to collect the dataset presented herein we simply query the private (prepublication) portion of our SNDB for all spectra observed between 1 January 2009 and 31 December 2018 for objects spectroscopically classified\footnote{We source spectroscopic classifications primarily from the Central Bureau of Electronic Telegrams (CBETs) and the International Astronomical Union Circulars (IAUCs).} as SNe~Ia.

\begin{figure}
	\centering
	\includegraphics[width=\columnwidth]{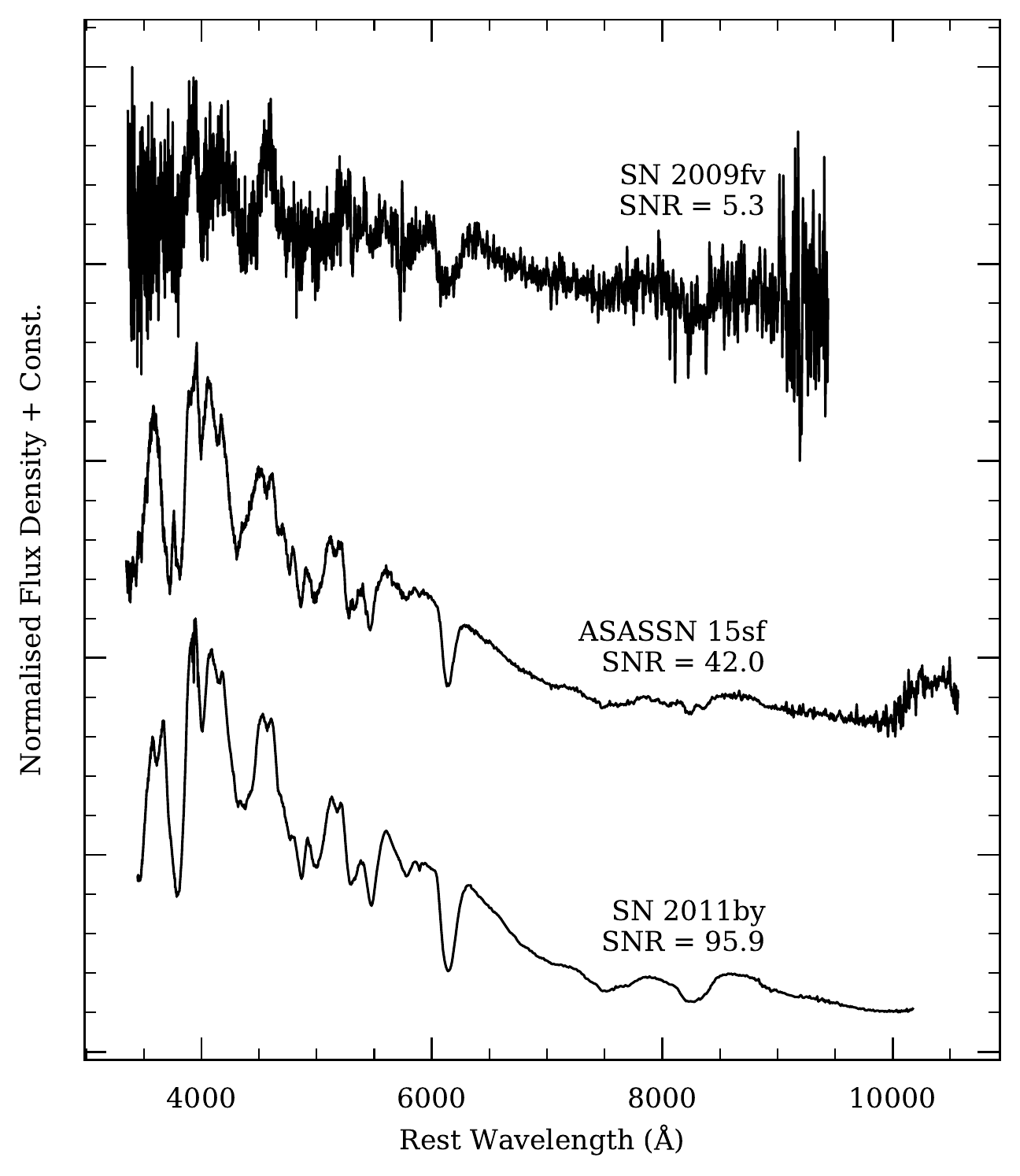}
	\caption{Representative SN Ia spectra from our sample showing low, medium, and high SNRs (progressing downward). The spectra have been deredshifted and normalised to a range of unity, and all are at $4 \pm 1$ days relative to to their SN's light-curve-determined time of maximum brightness.\label{fig:SNR-example}}
\end{figure}

This results in 744 matches, which we then filter by (i) selecting only those spectra with an average signal-to-noise ratio (SNR) greater than 5 pixel$^{-1}$ (yielding 714 matches above this quality threshold) and (ii) retaining only those with a wavelength coverage of at least 3700--7000~\AA\ (yielding 648 matches with sufficient spectral coverage for our subsequent analyses). Finally, we remove several of the remaining spectra, including any that are from SNe discovered earlier than 1 January 2008 (to avoid presenting only late-time spectra of an object at the early end of our selection range), to obtain the aforementioned set of 637 spectra. Following publication, all previously unpublished spectra will be transferred to the publicly accessible portion of the SNDB. We list basic SN-level information in Table~\ref{tab:SN-information} and spectrum-level information in Table~\ref{tab:spectra-information}, with many of the properties sourced from the Transient Name Server (TNS)\footnote{\url{https://wis-tns.weizmann.ac.il}} or the NASA/IPAC Extragalactic Database (NED)\footnote{The NASA/IPAC Extragalactic Database (NED) is operated by the Jet Propulsion Laboratory, California Institute of Technology, under contract with the National Aeronautics and Space Administration (NASA).}. Representative SN Ia spectra from our sample showing low, medium, and high SNRs are given in Figure~\ref{fig:SNR-example}. The SNR of the central spectrum in the figure is similar to the mean SNR for our entire sample (as discussed in Section~\ref{ssec:sample-characteristics}), and is thus indicative of the high quality of the spectra presented herein.

\begin{table*}
\caption{SN~Ia spectral information.\label{tab:spectra-information}}
\begin{tabular}{lcrcccrccrc}
\hline
\hline
\multicolumn{1}{l}{SN} & \multicolumn{1}{c}{UT Date$^a$} & \multicolumn{1}{c}{$t_{\rm LC}$$^b$} & \multicolumn{1}{c}{Instr.$^c$} & \multicolumn{1}{c}{Wavelength} & \multicolumn{1}{c}{Res.$^d$} & \multicolumn{1}{c}{P.A.$^e$} & \multicolumn{1}{c}{Airmass$^f$} & \multicolumn{1}{c}{Exposure} & \multicolumn{1}{c}{SNR} & \multicolumn{1}{c}{Reference$^{g}$}\\
\multicolumn{1}{l}{Name} & \multicolumn{1}{c}{(Y-M-D)} & \multicolumn{1}{c}{} & \multicolumn{1}{c}{} & \multicolumn{1}{c}{Range (\AA)} & \multicolumn{1}{c}{(\AA)} & \multicolumn{1}{c}{($^\circ$)} & \multicolumn{1}{c}{} & \multicolumn{1}{c}{Time (s)} & \multicolumn{1}{c}{} & \multicolumn{1}{c}{}\\
\hline
 SN 2008hm &  2008$-$12$-$31.311 &  $26.5$ &  1 &  3452$-$10700 &  4.3/10.5 &  $110.7$ &  $1.12$ &  1800 &  $31.7$ &  ... \\
 SN 2008hv &  2008$-$12$-$31.378 &  $14.3$ &  1 &  3452$-$10700 &  4.7/11.9 &  $138.8$ &  $1.32$ &  1200 &  $61.1$ &  ... \\
 SN 2008hy &  2009$-$01$-$05.155 &  $32.9$ &  1 &  3400$-$10700 &   4.9/9.8 &   $35.7$ &  $1.31$ &  1200 &  $24.9$ &  ... \\
  SN 2009D &  2009$-$01$-$05.184 &  $-5.6$ &  1 &  3390$-$10700 &  5.0/12.2 &  $161.1$ &  $1.86$ &  1200 &   $8.3$ &  ... \\
  SN 2009Y &  2009$-$02$-$19.665 &   $5.7$ &  2 &   3270$-$9270 &   2.1/6.7 &  $183.0$ &  $1.29$ &   180 &  $95.1$ &  ... \\
  SN 2009Y &  2009$-$03$-$29.532 &  $43.2$ &  1 &  3410$-$10100 &  5.3/11.3 &  $203.8$ &  $2.32$ &  1500 &   $9.6$ &  ... \\
  SN 2009Y &  2009$-$04$-$18.416 &  $62.9$ &  1 &   3454$-$9900 &  4.3/10.5 &  $181.7$ &  $1.76$ &  1800 &  $31.8$ &  ... \\
  SN 2009V &  2009$-$02$-$19.605 &     ... &  2 &   3388$-$9270 &   4.5/5.9 &   $95.0$ &  $1.68$ &   450 &   $7.5$ &  ... \\
 SN 2009ae &  2009$-$02$-$19.677 &     ... &  2 &   3270$-$9270 &   4.5/5.4 &   $85.0$ &  $1.02$ &   300 &  $39.6$ &  ... \\
 SN 2009an &  2009$-$03$-$29.507 &  $21.1$ &  1 &  3410$-$10100 &  4.5/12.1 &  $107.2$ &  $1.42$ &  1500 &  $34.1$ &  ... \\
\hline
\multicolumn{11}{p{17cm}}{Abridged table of SN~Ia spectral information (the full table is available as online supplementary material).}\\
\multicolumn{11}{p{17cm}}{$^{a}$Each UT date is specified for the temporal midpoint of the associated observation.}\\
\multicolumn{11}{p{17cm}}{$^{b}$Phases are in rest-frame days as computed from the appropriate redshift and photometry references from Table~\ref{tab:SN-information}.}\\
\multicolumn{11}{p{17cm}}{$^{c}$Instruments (Instr.) are as follows: (1) Kast (Shane 3~m) and (2) LRIS (Keck-I 10~m).}\\
\multicolumn{11}{p{17cm}}{$^{d}$Spectral resolution (Res.) are for the blue and red components, respectively. See Section 2 of S12a for more information.}\\
\multicolumn{11}{p{17cm}}{$^{e}$Observed slit position angle (P.A.) for each observation.}\\
\multicolumn{11}{p{17cm}}{$^{f}$Each airmass is specified for the temporal midpoint of the associated observation.}\\
\multicolumn{11}{p{17cm}}{$^{g}$References to previous publications including the noted spectra are as follows: (1) \citet{SN2009dc}, (2) \citet{SN2009ig}, (3) \citet{Iax}, (4) \citet{bsnipV}, (5) \citet{Mazzali2015}, (6) \citet{SN2012cg}, (7) \citet{SN2012fr}, (8) \citet{SN2013dy}, (9) \citet{SN2013dy-Pan}, (10) \citet{SN2014dt}, (11) \citet{Iax-late-spec}, (12) \citet{SN2016coj}, and (13) Xuhui et al. (2019, in prep.).}
\end{tabular}
\end{table*}

\subsection{Observations}
\label{sec:observations}

The vast majority of the spectra in our dataset (579/637) were obtained using the Kast double spectrograph \citep{Kast} mounted on the Shane 3~m telescope at the Lick Observatory. The remaining observations (58/637) were made with the Low Resolution Imaging Spectrometer \citep[LRIS;][]{LRIS} at the W. M. Keck Observatory. The seeing at these locations averages $\sim 2^{\prime \prime}$ and $\sim 1^{\prime \prime}$, respectively. Most spectra presented here were obtained with the long slit at or near the parallactic angle so as to reduce the differential light loss caused by atmospheric dispersion \citep{Filippenko1982}; however, this was not necessary with LRIS, as it is equipped with an atmospheric dispersion corrector. The specific details of our observing strategy are thoroughly documented by S12a, so here we mention only relevant changes to the aforementioned instruments.

On 18 September 2016, the Kast red-side CCD was replaced with a Hamamatsu $1024\times4096$ pixel device with 15~$\mu$m pixels, yielding a spatial scale of $0\overset{\prime \prime}{.}43$ pixel$^{-1}$. Compared to the previous red-side CCD, the new detector features significantly reduced readout noise and better quantum efficiency for wavelengths greater than 5000~\AA. Most (483/579) Kast spectra presented herein were taken prior to this upgrade.

In May and June of 2009, the LRIS red-channel CCD was replaced with a mosaic of two 2k $\times$ 4k pixel Lawrence Berkeley National Lab (LBNL) CCDs with a spatial scale of $0\overset{\prime \prime}{.}135$ pixel$^{-1}$. The mosaic features smaller pixels and higher quantum efficiency in the red than the original CCD \citep{LRIS_update}. Nearly all (52/58 since 1 July 2009) LRIS spectra were taken using this upgraded configuration.

\subsection{Data Reduction}
\label{sec:data_reduction}

An important attribute of our sample is the consistency with which the data have been reduced. Regardless of instrument, the same general procedures are followed for all spectral reductions, and just five individuals are responsible for reducing the majority ($> 88\%$) of our dataset. In the following paragraph, we briefly summarise the principal steps in our reduction strategy (see S12a for a more comprehensive discussion), which are implemented using {\tt IRAF}\footnote{IRAF is distributed by the National Optical Astronomy Observatory, which is operated by AURA, Inc., under a cooperative agreement with the U.S. National Science Foundation (NSF).} routines and publicly available {\tt Python} and {\tt IDL} programs\footnote{Kast and LRIS data are currently reduced with {\tt KastShiv} \citep{Shivvers16} and {\tt LPipe} \citep{LPipe}, respectively. Prior to October 2016, a number of LRIS spectra were reduced with purpose-built routines from the Carnegie {\tt Python} ({\tt CarPy}) Distribution \citep{carpy1,carpy2}.}.

First, standard preparation steps including bias removal, cosmic ray rejection, and flat-field correction are performed. Following extraction, one-dimensional spectra are wavelength-calibrated using comparison-lamp spectra typically taken in the afternoon prior to each observing run. The spectra are then flux-calibrated using spectra (taken during each observing run with the appropriate instrumental setup) of bright spectrophotometric standard stars at similar airmasses. Finally, atmospheric (telluric) absorption features are removed and overlapping (i.e., red- and blue-side spectra from Kast or LRIS) are combined by scaling one so that it matches the other\footnote{For Kast spectra, the blue side is scaled to match the red side, while for LRIS spectra, whichever side shows the lower transmission level is scaled upward.} over the common wavelength range. We consider spectra at this stage to be ``science ready.''

\section{Classification}
\label{sec:classification}

Optical spectra are widely used to classify SNe as belonging to one of several distinct types, and possibly subtypes \citep[e.g.,][]{Filippenko1997}. We perform such classification in an automated fashion using the SuperNova IDentification code \citep[{\tt SNID},][]{snid} with tightly controlled tolerances. {\tt SNID} classifies SNe by cross-correlating an input spectrum against a large library of template spectra \citep{Tonry1979}. In the following sections we detail our spectral classification procedure, present results, and discuss verifications of these results.

\subsection{{\tt SNID} Classification Procedure}
\label{sec:snid_procedure}

Using a classification scheme similar to that employed by S12a, we attempt to determine the type, subtype, redshift, and age from each spectrum in our sample via consecutive {\tt SNID} runs that adhere to the specifications outlined in the following sections.

\subsubsection{{\tt SNID} Type}
\label{sec:snid_type}

We first attempt to determine the type of a SN from its spectrum by executing a {\tt SNID} run and requiring an \emph{r}lap\footnote{The \emph{r}lap is a measure of quality used by {\tt SNID} --- higher values correspond to classifications that are more trustworthy.} value of at least 10. If the host-galaxy redshift of the SN is listed in Table~\ref{tab:SN-information}, then we force {\tt SNID} to use this redshift by invoking the \emph{forcez} keyword --- otherwise {\tt SNID} will attempt to find the redshift simultaneously. In order for type determination to be considered successful, we require that the fraction of ``good''\footnote{In SNID, a template is graded ``good'' when its strongest correlation with the input spectrum occurs at a redshift that differs by less than 0.02 from the forced (or simultaneously fit) redshift of the input spectrum.} correlations corresponding to the proposed type be $>50$\% and that the best-matching ``good'' template be of the same type. If no type is determined by this approach, we relax the minimum \emph{r}lap value to 5 and repeat the procedure. If a type is determined at this stage, we proceed to subtype determination.

\subsubsection{{\tt SNID} Subtype}
\label{sec:snid_subtype}

In the subtype-determination run, we again force {\tt SNID} to use the redshift of the SN if it is available (and find it simultaneously otherwise). We also force {\tt SNID} to use only templates that match the previously found type. Again, we attempt a {\tt SNID} run with a minimum \emph{r}lap value of 10, and relax this to 5 if the first run is unsuccessful. In the case of subtype determination, success is achieved if the fraction of ``good'' correlations corresponding to a subtype is $>50$\% and the best-matching ``good'' template is of the same subtype.

\subsubsection{{\tt SNID} Redshift}
\label{sec:snid_redshift}

We use {\tt SNID} to determine the redshift from a spectrum by executing a {\tt SNID} run that requires all templates to be of the subtype found previously (or type, if the subtype was not successfully determined). We use no external redshift information, even if it appears in Table~\ref{tab:SN-information}, but we do restrict the range of template redshifts to lie within $0 < z < 0.3$. We calculate the redshift as the median of all ``good'' template redshift values, and the redshift uncertainty is taken to be the standard deviation of these values. If the redshift and subtype are determined, then we attempt to find the rest-frame phase relative to maximum light (henceforth referred to as ``age'') from the spectrum.

\subsubsection{{\tt SNID} Age}
\label{sec:snid_age}

We attempt to determine the age of a SN spectrum by executing a {\tt SNID} run that uses only templates of the subtype determined previously and that requires {\tt SNID} to use the known redshift, or the redshift determined previously if it was not known. The age (henceforth, $t_{\rm SNID}$) is calculated as the median of only the ``good'' template ages that have an \emph{r}lap value of at least 75\% of the largest achieved \emph{r}lap value. The age uncertainty is the standard deviation of these ages. Furthermore, we require that the age uncertainty be less than the larger of 4 days or 20\% of the determined age.

\subsection{Classification Results and Verifications}
\label{sec:classification-results}

Of the 637 spectra selected for characterisation, our \texttt{SNID} routine successfully determines the type in 608 instances, the subtype in 506, the redshift in 605, and the age in 406. We present the results derived from performing our \texttt{SNID} classification procedure in Table~\ref{tab:snid-results}, and we discuss and examine them in the following subsections.

\begin{table*}
\caption{{\tt SNID} classification results.\label{tab:snid-results}}
\begin{tabular}{l|cccr|lcccr}
\hline
\hline
& \multicolumn{4}{c|}{Classification Results} & \multicolumn{5}{c}{Best-Matching {\tt SNID} Template}\\
\multicolumn{1}{l|}{SN} & \multicolumn{1}{c}{Type} & \multicolumn{1}{c}{Subtype} & \multicolumn{1}{c}{$z_{\rm SNID}$} & \multicolumn{1}{c|}{$t_{\rm SNID}$$^{a}$} & \multicolumn{1}{c}{Name} & \multicolumn{1}{c}{Subtype} & \multicolumn{1}{c}{\emph{r}lap} & \multicolumn{1}{c}{$z$} & \multicolumn{1}{c}{$t^a$}\\
\hline
 SN 2008hm &  Ia &      ... &  0.0195 $\pm$ 0.0046 &              ... &   sn99aa &   Ia-99aa &  17.5 &  0.0321 &  $34.1$ \\
 SN 2008hv &  Ia &  Ia-norm &  0.0114 $\pm$ 0.0037 &   $18.9 \pm ...$ &   sn04ey &   Ia-norm &  31.2 &  0.0114 &  $18.9$ \\
 SN 2008hy &  Ia &      ... &  0.0076 $\pm$ 0.0038 &              ... &    sn91T &    Ia-91T &  23.4 &  0.0054 &  $46.6$ \\
  SN 2009D &  Ia &  Ia-norm &  0.0214 $\pm$ 0.0063 &  $-10.6 \pm 2.8$ &    sn90N &   Ia-norm &  16.7 &  0.0301 &  $-6.4$ \\
  SN 2009Y &  Ia &  Ia-norm &  0.0014 $\pm$ 0.0060 &    $5.9 \pm 3.2$ &   sn02bo &   Ia-norm &  14.0 &  0.0041 &   $5.5$ \\
  SN 2009Y &  Ia &  Ia-norm &  0.0085 $\pm$ 0.0023 &   $44.5 \pm 5.7$ &   sn02bo &   Ia-norm &  14.0 &  0.0070 &  $44.5$ \\
  SN 2009Y &  Ia &  Ia-norm &  0.0116 $\pm$ 0.0031 &   $72.5 \pm 9.7$ &   sn02bo &   Ia-norm &  14.1 &  0.0091 &  $72.5$ \\
  SN 2009V &  Ia &  Ia-norm &  0.0933 $\pm$ 0.0044 &   $11.2 \pm 0.5$ &   sn94ae &   Ia-norm &  15.6 &  0.0938 &  $11.3$ \\
 SN 2009ae &  Ia &  Ia-norm &  0.0307 $\pm$ 0.0043 &   $18.2 \pm 3.4$ &   sn02bo &   Ia-norm &  17.2 &  0.0345 &  $17.8$ \\
 SN 2009an &  Ia &  Ia-norm &  0.0078 $\pm$ 0.0030 &              ... &   sn02eu &   Ia-norm &  14.5 &  0.0065 &  $33.4$ \\
\hline
\multicolumn{10}{p{13cm}}{Abridged table of {\tt SNID} classifications (the full table is available as online supplementary material).}\\
\multicolumn{10}{p{13cm}}{$^{a}$Spectral ages (phases) are in rest-frame days relative to the time of the associated SN's maximum brightness. Age uncertainties marked with ``...'' correspond to cases where only one template was a ``good'' match.}
\end{tabular}
\end{table*}

\subsubsection{Types and Subtypes}
\label{sssec:type-checks}

To study the robustness of our \texttt{SNID}-determined types and subtypes, we look for distinctions we can draw between spectra that were successfully classified versus those that were not. In particular, we investigate whether there is a significant difference between success and failure that is codified by (i) the average SNR of a spectrum, or (ii) the phase in a SN's temporal evolution during which that spectrum was observed.

The median SNR of the spectra for which \texttt{SNID} successfully determines a type (subtype) is 32.7 pixel$^{-1}$ (33.4 pixel$^{-1}$), while for those where it failed the median is 14.3 pixel$^{-1}$ (26.4 pixel$^{-1}$). For the case of determining the type, this presents a compelling argument --- spectra for which the type is classified are generally of higher quality (as assessed by the SNR) than those that are not. Although the gap in median SNR between the successful and failed subsets is notably less pronounced for the case of determining the subtype, we must make a concession for the fact (as stated in Section~\ref{sec:snid_type}) that the entire population, for which an attempt is made to determine subtype, is drawn \emph{only} from those where the type has been successfully determined (and hence whose aggregate SNR is higher, as discussed above). With this important caveat noted, it would appear that the gap in SNR between successful and failed subtype classifications is indeed meaningful --- those SNe for which the subtype is not successfully determined have a median SNR that is $\sim 9$ times below that of the entire population, relative to the median SNR for those for which the subtype is determined.

Next, we examine how the difference in rest-frame days between when a spectrum was observed and when the SN in that spectrum reached maximum brightness as determined from its light curve (i.e., the phase) may influence \texttt{SNID}'s success rate with regard to (sub)type classification. We find that the median phase in cases where \texttt{SNID} successfully identifies a type (subtype) is 19.4 days (16.5 days), while in cases where it fails the median is 65.3 days (40.7 days). Owing to the much sparser coverage of \texttt{SNID} spectral templates at late phases (see, e.g., Figure 6 of S12a), it makes sense for the failure rate to be larger for spectra at late phases. In addition, spectra at earlier phases tend to have higher SNRs than do those at later phases\footnote{If we divide our sample into two groups based on phase ($<20$ days, $>20$ days), the median SNR of the early-time subset is 57.6 pixel$^{-1}$, while for the late-time subset it is 32.4 pixel$^{-1}$.} because SNe~Ia fade throughout their post-maximum evolution. As we have seen above, the SNR of a spectrum plays a substantive role the outcome of (sub)type classification. We find it reasonable, then, that the median phase is earlier for successes than it is for failures. Furthermore, while the caveat from the preceding paragraph regarding the population for which subtype-determination is attempted is still relevant, it is similarly overcome --- the difference between the median phase of those for which the subtype is not successfully determined (40.7 days) and that of the entire population (19.4 days) is $\sim 7$ times larger than the associated difference for those whose subtype is determined (16.5 days).

\subsubsection{Redshifts}
\label{sssec:redshift-checks}

\begin{figure}
	\centering
	\includegraphics[width=\columnwidth]{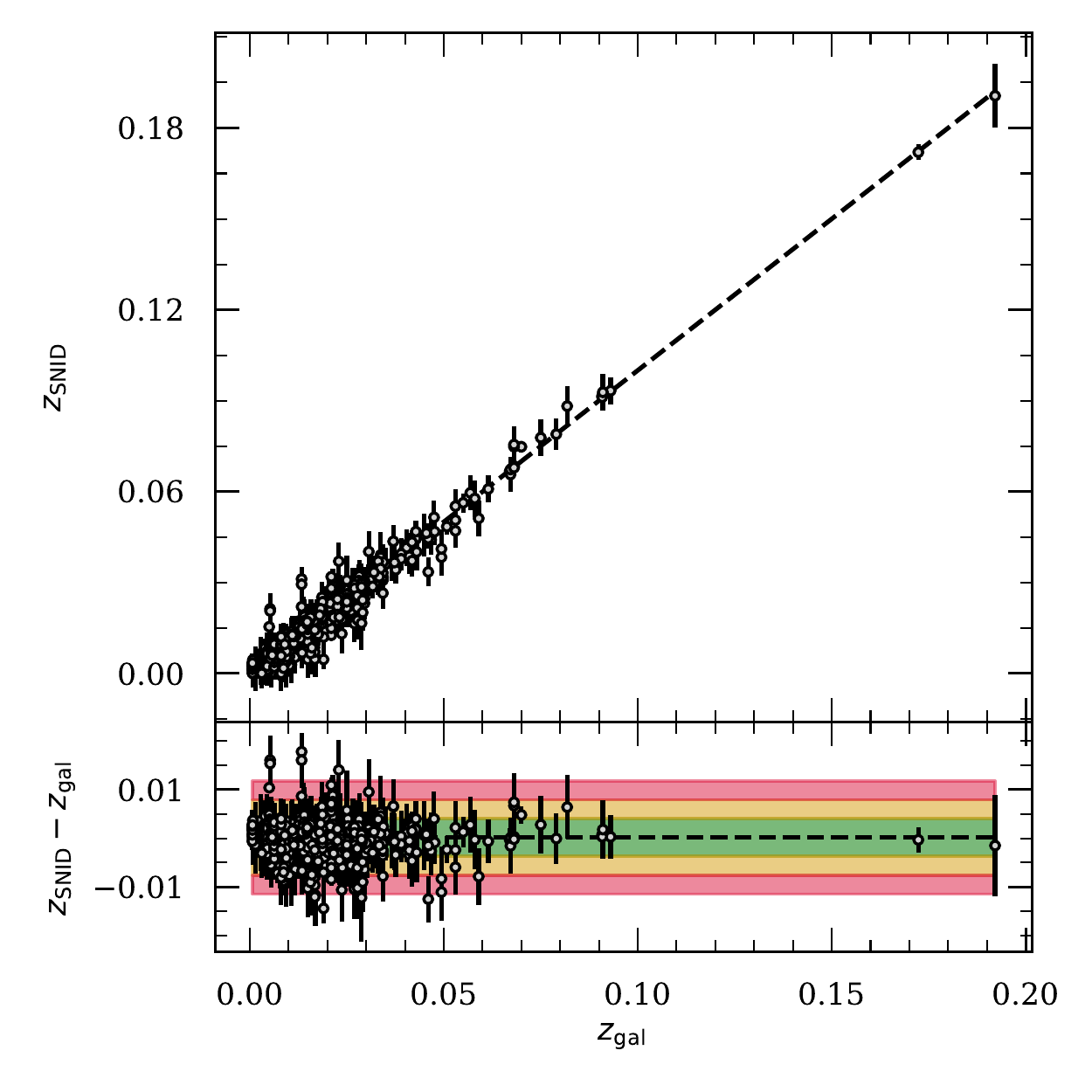}
	\caption{SNID-determined redshifts versus host-galaxy redshifts, with residuals in the lower panel. The dashed line in the top panel shows the one-to-one correspondence for $z_\mathrm{gal}$, and in the bottom panel it indicates the median residual. The green, yellow, and red regions in the lower panel correspond to the $1\sigma$, $2\sigma$, and $3\sigma$ bounds about the median residual, respectively. We note that the typical uncertainties for $z_{\rm gal}$ (which are omitted from the figure) are $\sim 1/4$ of those for $z_{\rm SNID}$.\label{fig:z_verif}}
\end{figure}

We investigate our {\tt SNID}-determined redshifts by comparing them to the corresponding host-galaxy redshifts, when they are available, as shown in Figure~\ref{fig:z_verif}. From the 563 spectra in our sample for which (i) {\tt SNID} determined a redshift, (ii) {\tt SNID} determined the spectrum was of a SN~Ia (independent of the subtype classification), and (iii) a redshift is listed in Table~\ref{tab:SN-information}, we find a median residual of 0.0002 with a standard deviation of 0.0039. Furthermore, we calculate the normalised median absolute deviation \citep{Ilbert}, defined as
\begin{equation}
	\sigma \equiv 1.48 \times \text{median} \left[\frac{|z_\text{SNID} - z_\text{gal}|}{1 + z_\text{gal}}\right],
\end{equation}
and find a value 0.003, similar to S12a who found 0.002 for their dataset. Of the spectra used for comparison, 446 have a redshift residual within one standard deviation of the median, 522 are within two standard deviations, and 553 are within three.

\subsubsection{Phases}
\label{sssec:phase-checks}

\begin{figure}
	\centering
	\includegraphics[width=\columnwidth]{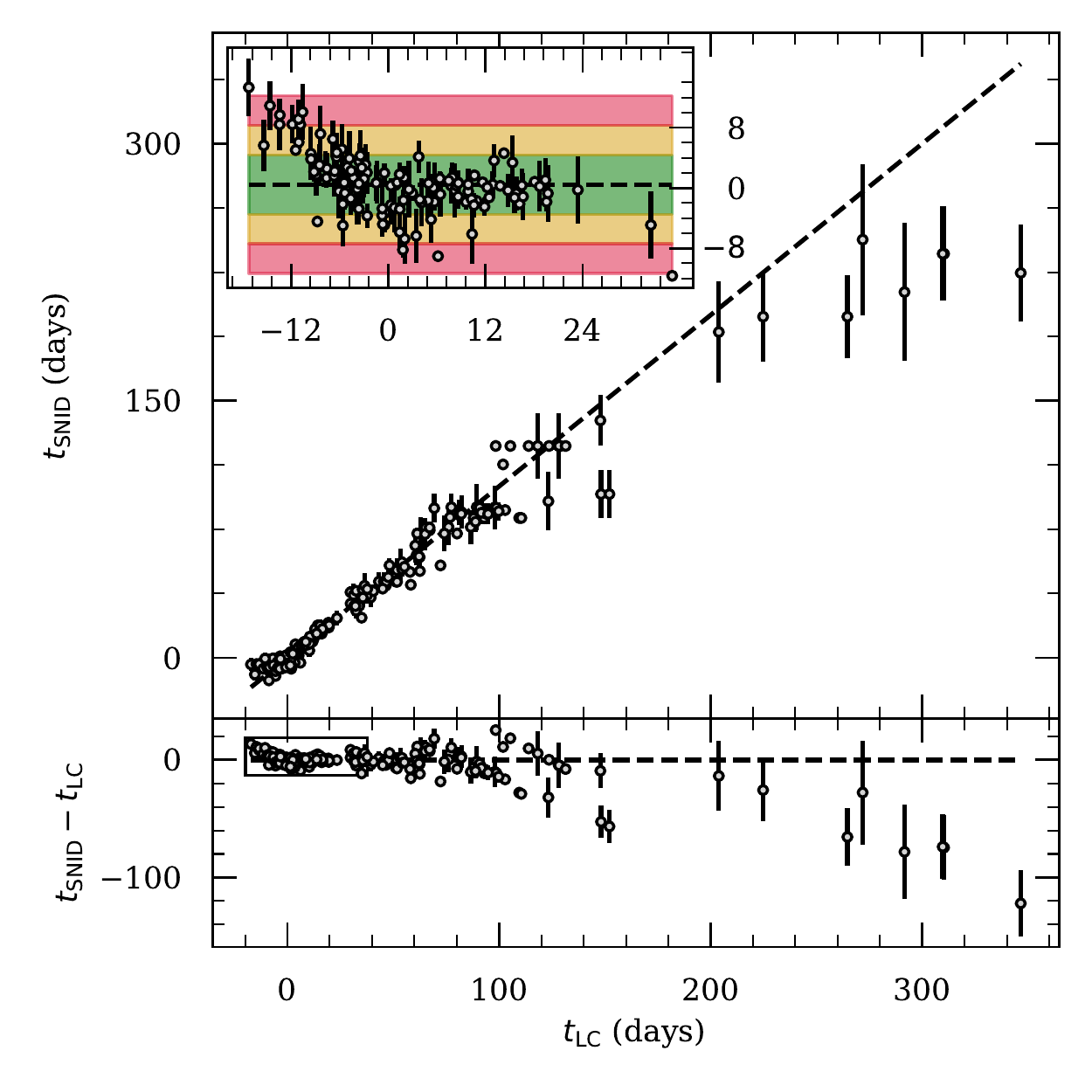}
	\caption{SNID-determined phases versus those derived from light-curve maxima and listed in Table~\ref{tab:spectra-information}. Residuals are shown in the lower panel. The dashed line in the top panel indicates the one-to-one correspondence for $t_\mathrm{LC}$, and in the bottom panel it shows the median residual. The inset panel displays the residuals from the subset of our sample for which $t_\mathrm{SNID} \leq 30$ days and $t_\mathrm{LC} \leq 50$ days, in addition to the conditions used to select the initial sample. The green, yellow, and red regions correspond to the $1\sigma$, $2\sigma$, and $3\sigma$ bounds about the median residual, respectively.\label{fig:t_verif}}
\end{figure}

Next we compare \texttt{SNID}-determined phases to those calculated (in rest-frame days) relative to light-curve-determined times of maxima (henceforth, $t_{\rm LC}$), when available (see Table~\ref{tab:spectra-information} for $t_{\rm LC}$ values and Table~\ref{tab:SN-information} for references on the times of maximum brightness used to compute them). We perform this comparison for all spectra with the requisite information which {\tt SNID} classified as belonging to a SN~Ia (for a total of 219 spectra), and the result is shown in Figure~\ref{fig:t_verif}. There is a rather tight correlation for $t_\mathrm{SNID} \lesssim 100$ days, but beyond this point the {\tt SNID}-determined ages systematically underestimate the true (i.e., light-curve-derived) phases. This is not unexpected given the dearth of template spectra available at late phases (as discussed in Section~\ref{sssec:type-checks}) and is consistent with the results of previous studies (e.g., Figure 7 of S12a).

If we further restrict the subset used for phase comparison to cover only the earlier, more rapidly evolving stages of spectroscopic evolution [namely, only those for which the (rest-frame) light-curve-determined phase is $<50$ days and the {\tt SNID}-determined phase is $<30$ days], we are left with 127 spectra. The median residual for this subset is $\sim 0.4$ days with a standard deviation of $\sim 3.9$ days. Of this subset, 95 spectra have a residual that lies within $1\sigma$ of the median, 116 are within $2\sigma$, and 125 are within $3\sigma$. We find that for very early phases ($t_\mathrm{LC} \lesssim -10$ days), {\tt SNID}-determined phases tend to be an overestimate (as can be seen in the inset panel of Figure~\ref{fig:t_verif}). As with {\tt SNID}'s tendency to underestimate the phase of late-time spectra, the dominant cause of the noted early-phase overestimate can be attributed to the paucity of template spectra at similar phases.

\subsection{Object Classifications}
\label{ssec:object-classifications}

Many of the SNe in our sample have multiple spectra, and therefore we must combine the classification information derived for each spectrum to obtain a final classification for each object. To determine the type of an object with multiple spectra, we choose the most frequently occurring type in that object's spectral classifications. In cases with a tie between two possible type characterisations, we use the type of the spectrum whose best-matching {\tt SNID} template has the larger \emph{r}lap value. To account for the uncertainty surrounding such classifications, we add a ``*'' to the type. We follow a similar procedure for determining the subtype of each object, except that in cases where there is a tie for the most frequent subtype, we do not classify the subtype. The final (sub)type derived from this methodology is listed for each SN in our sample in Table~\ref{tab:SN-information}. Altogether, 242 objects are unambiguously classified as SNe~Ia and one is given the classification of ``Ia*''. The remaining four objects are discussed in the following section.

\subsection{Objects Not Classified as SNe~Ia}

There are four objects (SN~2009eq, LSQ~12fhe, SN~2013gh, and SN~2013fw) in our dataset for which the aforementioned classification method either fails to classify the object at all, or classifies it as something other than a SN~Ia. We examine and briefly discuss each of these objects below.

\subsubsection{SN~2009eq}

Of the three spectra of SN~2009eq included in our dataset, two (taken 3 d and 20 d after our first spectrum) are classified as belonging to a SN~Ic, and the remaining one (our first observation of the object) is not successfully classified at all. After visual inspection of the three spectra by multiple coauthors, we override the {\tt SNID}-determined type in favour of ``Ia*'' --- the spectra appear to be consistent with that of a SN~Ia, and particularly a SN~1991bg-like \citep{91bg-Filippenko,91bg-Leibundgut} object evolving within one month of maximum brightness. However, given that our {\tt SNID}-based classification scheme does not come to the same conclusion, we cannot unambiguously give a ``Ia'' classification \emph{from our dataset alone}. It is also worth noting that our determination that SN~2009eq is a SN~1991bg-like object is consistent with its initial classification \citep{2009eq}.

\subsubsection{LSQ~12fhe}

Our classification scheme deems the single spectrum of LSQ~12fhe in our dataset to be of a SN~Ic, contradicting the object's initial classification as a SN~Ia of the SN~1991T-like \citep{91T-Filippenko,91T-Phillips} subtype \citep{LSQ12fhe}. Looking more closely at our {\tt SNID} classification, we see that the SN~Ic classification was favoured by just one more template than for a SN~Ia. After visual inspection by multiple coauthors, we reach a consensus that the object is definitely a SN~Ia, and most likely of the SN~1991T-like subtype (consistent with the initial classification). Accordingly, we override our {\tt SNID}-determined type to be ``Ia*'' --- this reflects its true classification but accounts for the fact that our classification scheme does not reach the correct conclusion.

\subsubsection{SN~2013gh}

SN~2013gh is covered by three spectra in our dataset (with light-curve-determined phases of $-12$, 70, and 392 days). The first spectrum is unambiguously determined to be of a SN~Ia (with an undetermined age), while the second is assigned as a SN~Ic (with an age of 1.5 days), and the third is undetermined (not surprising given {\tt SNID}'s lack of late-phase templates, as previously discussed). In light of (i) the visually obvious SN~Ia determination from the first spectrum, (ii) the completely incorrect {\tt SNID}-determined phase of the second spectrum, and (iii) the multiple-coauthor consensus that the second spectrum is consistent with that of a SN~Ia at the appropriate phase, we again override our {\tt SNID}-determined type in favour of ``Ia*''.

\subsubsection{SN~2013fw}

The single spectrum of SN~2013fw in our dataset is at a very late phase ($> 300$ days), and thus it is unsurprising, given {\tt SNID}'s lack of suitable templates (as discussed in Section~\ref{sssec:type-checks}), that our classification scheme does not succeed. We thus defer to the existing object classification \citep{2013fw_CBET}, and assign its type as ``Ia*'' --- it is a SN~Ia but we cannot conclusively confirm or refute the classification using our dataset alone.

\section{Results}
\label{sec:results}

In this section we present and study our low-redshift SN~Ia spectral dataset derived from observations totaling more than 275~hr of telescope time. Of our initial selection (from Section~\ref{ssec:selection}), 242 objects (covered by a total of 626 spectra) are unambiguously classified as SNe~Ia by the methodology described in the preceding section. In the discussion that follows, we consider only this selection of spectra. We provide plots and file access for all spectra described in this work electronically via our SNDB.

\subsection{Sample Characteristics}
\label{ssec:sample-characteristics}

\begin{figure*}
	\centering
	\includegraphics[width=\textwidth]{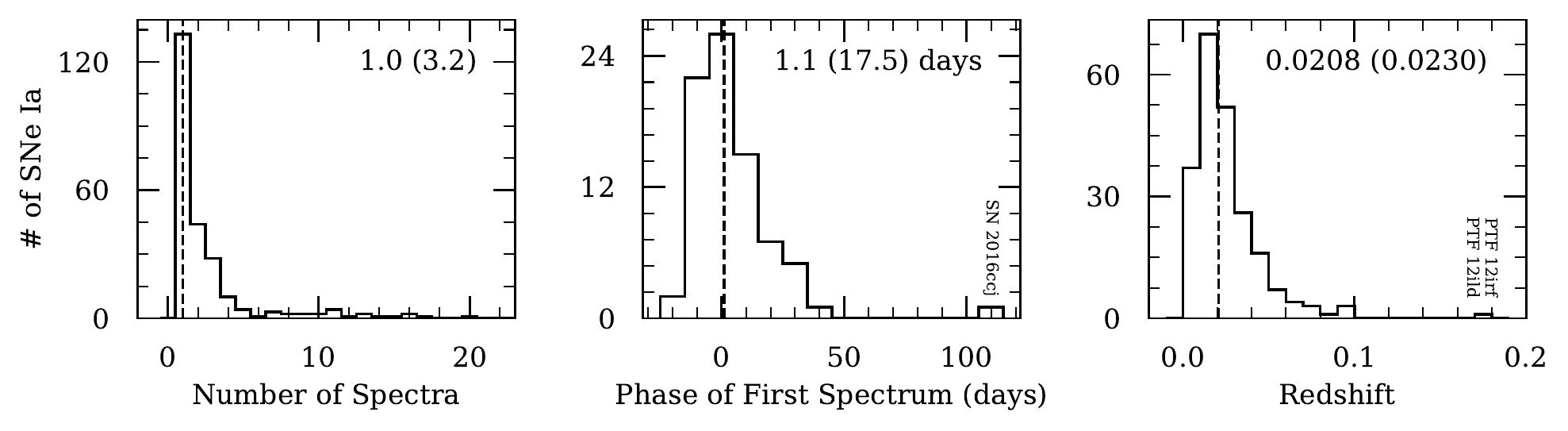}
	\caption{Distributions of SN-level parameters, with the associated median (standard deviation) values included. The left panel is the number of spectra, centre is the light-curve-determined rest-frame phase of the first observed spectrum, and right is redshift. The SNe responsible for the outlying bins in the centre and right panels are labeled.\label{fig:sn_chars}}
\end{figure*}

Our dataset averages 2.6 spectra per SN~Ia (with a median of 1), similar to the $\sim 2.2$ spectra per object S12a found for their dataset and reflective of BSNIP's emphasis on maximising the number of objects studied spectroscopically rather than the number of spectra per object. SN~2016coj has the most spectra of any object in our sample with 20, followed by SN~2011fe with 17. Figure~\ref{fig:sn_chars} shows the full distribution of the number of spectra per SN~Ia. Of the 242 SNe~Ia in our sample, 109 are covered by at least two spectra. For the 79 SNe in our sample that have a light-curve-determined time of maximum brightness (as noted in Table~\ref{tab:SN-information}), we find a median (rest-frame) phase of the first spectrum of 1.1 days, as shown in the centre panel of Figure~\ref{fig:sn_chars}. Of this subsample with phase information, 38 SNe have a spectrum observed before the time of maximum brightness and 69 have one within 20 days of maximum. We show the redshift distribution of the objects in our sample in the right panel of Figure~\ref{fig:sn_chars}. Aside from two SNe with redshifts of near (but below) 0.2, all have $z < 0.1$ and 201 (of the 221 with a redshift listed in Table~\ref{tab:SN-information}) have $z \leq 0.05$. We find a median redshift of 0.0208 for the full sample, and for the 184 SNe with $z \geq 0.01$ (i.e., within the Hubble flow) we find a median of 0.0230.

\begin{figure*}
	\centering
	\includegraphics[width=\textwidth]{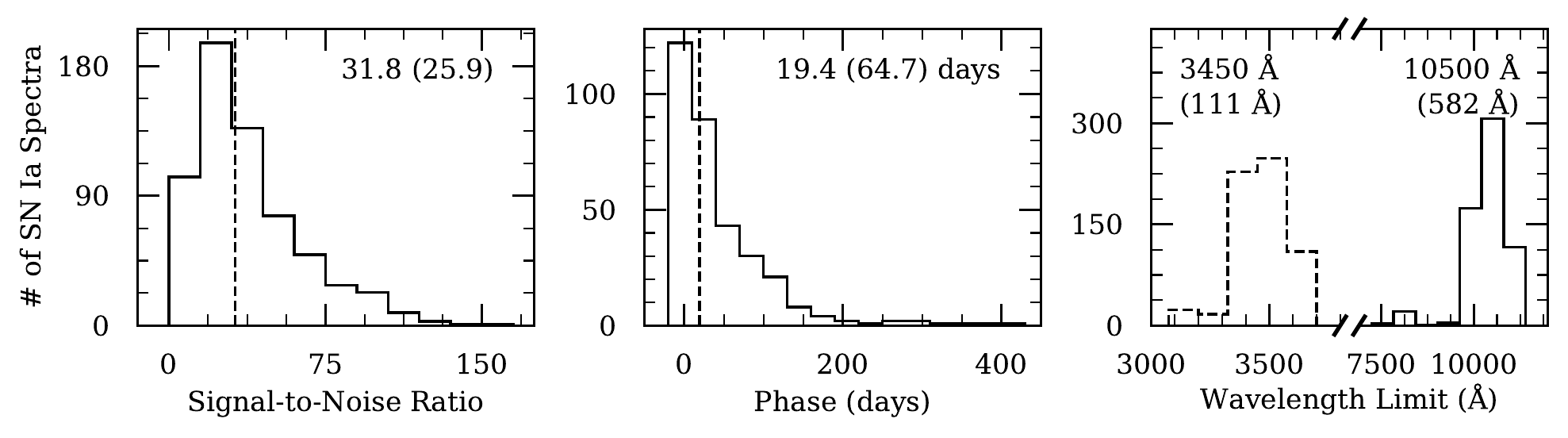}
	\caption{Distributions of spectrum-level parameters, with the associated median (standard deviation) values included. The left panel shows the SNR, centre is the light-curve-determined rest-frame phase, and right is the blue and red wavelength limits.\label{fig:spectra_chars}}
\end{figure*}

We show the distribution of average SNRs for the spectra in our dataset in the left panel of Figure~\ref{fig:spectra_chars}. The median is 31.8 pixel$^{-1}$ (with a mean of 38.3 pixel$^{-1}$), and 574/626 spectra have SNR $> 10$ pixel$^{-1}$. By design (see Section~\ref{ssec:selection}), we find a minimum SNR of $\sim 5$ pixel$^{-1}$. As shown in the centre panel of Figure~\ref{fig:spectra_chars}, we find the median (light-curve-determined rest-frame) phase for the spectra with such information to be 19.4 days. The spectrum with the earliest phase belongs to SN~2011fe at $-17.2$ days, followed by two spectra of SN~2012cg with phases of $-16.4$ days and $-15.4$ days. The spectrum with the latest phase belongs to SN~2013dy at 422 days, followed by one from SN~2011fe at 379 days. Our dataset includes 15 spectra at phases of at least 160 days. We find that 168 of the 328 spectra in our sample which have light-curve-determined phases correspond to earlier than 20 days in the post-maximum evolution of their SN. The distributions of the wavelengths of the blue and red ends of our spectra are shown in the right panel of Figure~\ref{fig:spectra_chars}. We find a median blue (red) wavelength limit of 3450~\AA\ (10,500~\AA), and 592 of our spectra cover at least 3700--9800~\AA.

\subsection{Early-time Spectra}

A number of prior SN~Ia analyses \citep[e.g.,][]{Riess1997-SFA,Folatelli-pEW,Foley2005-SFA,Branch2006,Garavini07,Wang2009,Blondin2012,bsnipII,Folatelli2013,Childress14,Zhao15} have studied SN~Ia optical spectra in terms of multiple ``features'' --- each typically a blend of many spectral transitions, but distinctive enough to be considered in aggregate as a single major absorption feature complex. Of principal interest are assessments of (i) the expansion velocities of such features, and (ii) quantities that probe the relative strengths of the features, often assessed through pseudo-equivalent width (pEW) measurements.

Providing a tracer of explosion kinetic energy, the expansion velocities of SN~Ia ejecta have been extensively studied --- especially during the characteristic decline through the near-maximum evolution \citep[e.g.,][]{Benetti2005,Wang2009}. \citet[][henceforth S12b]{bsnipII} find velocities within a few days of maximum brightness that are consistent with the notion that SN~Ia ejecta are layered --- features of \ion{O}{i}, \ion{Si}{ii}, and \ion{S}{ii} tend to have lower velocities (and are thus found in the inner, more slowly expanding layers), while those of \ion{Ca}{ii} have the highest velocities (and are therefore associated with the outer, more rapidly expanding layers). These findings are consistent with our own (see Section~\ref{sssec:exp-velocity}). Together with probes of feature strength (e.g., pEW measurements), expansion velocities can be used to quantify the degree of homogeneity between spectra of different SNe~Ia (and hence SNe themselves) at similar epochs, as well as describe the expected temporal evolution of spectral features \citep{Folatelli-pEW}. Feature-strength measurements from SN~Ia spectra are further prized for the prospect that they might correlate with luminosity \citep[e.g.,][]{Nugent1995,bsnipIII}.

Following S12b, we measure the expansion velocities, pEWs, and fluxes at the endpoints of nine features in the spectra from our sample which have a light-curve-determined rest-frame phase of $< 20$ days\footnote{Two pairs of the selected features (\ion{Si}{ii} $\uplambda 4000$, \ion{Mg}{ii}; and \ion{S}{ii} ``W'', \ion{Si}{ii} $\uplambda 5972$) become significantly blended at the late end of this range, so we therefore only measure these features for $t_{\rm LC} \leq 10$ days.}. While some studies consider high-velocity and photospheric components for certain features \citep[typically by fitting a series of Gaussians to the absorption profile; e.g.,][]{Silverman-highv,Pan2015,Zhao2016}, we do not draw such a distinction in the following analysis (so as to remain consistent with the methodology of S12b). Our selected features, each labeled by the ion or spectral transition line most dominant in the absorption, are listed in Table~\ref{tab:spectral-features} along with their rest wavelengths.

\begin{table}
\caption{Spectral features.\label{tab:spectral-features}}
\begin{tabular}{lcccc}
\hline
Feature & Rest & Blue & Red \\
 & Wavelength (\AA) & Boundary (\AA) & Boundary (\AA) \\
\hline
\ion{Ca}{ii} H\&K & 3945.28 & 3400--3800 & 3800--4100 \\
\ion{Si}{ii} $\uplambda 4000$ & 4129.73 & 3850--4000 & 4000--4150 \\
\ion{Mg}{ii} & $...^a$ & 4000--4150 & 4350--4700 \\
\ion{Fe}{ii} & $...^a$ & 4350--4700 & 5050--5550 \\
\ion{S}{ii} ``W'' & 5624.32 & 5100--5300 & 5450--5700 \\
\ion{Si}{ii} $\uplambda 5972$ & 5971.85 & 5400--5700 & 5750--6000 \\
\ion{Si}{ii} $\uplambda 6355$ & 6355.21 & 5750--6060 & 6200--6600 \\
\ion{O}{i} triplet & 7773.37 & 6800--7450 & 7600--8000 \\
\ion{Ca}{ii} near-IR triplet & 8578.75 & 7500--8100 & 8200--8900 \\
\hline
\end{tabular}
\textbf{Note:} Spectral features and boundaries, as adapted from S12b.\\
$^a$A single reference wavelength is not useful for this feature because it is a blend of too many spectral lines. Hence, we do not compute expansion velocities for this feature.
\end{table}

\begin{figure*}
	\centering
	\includegraphics[width=\textwidth]{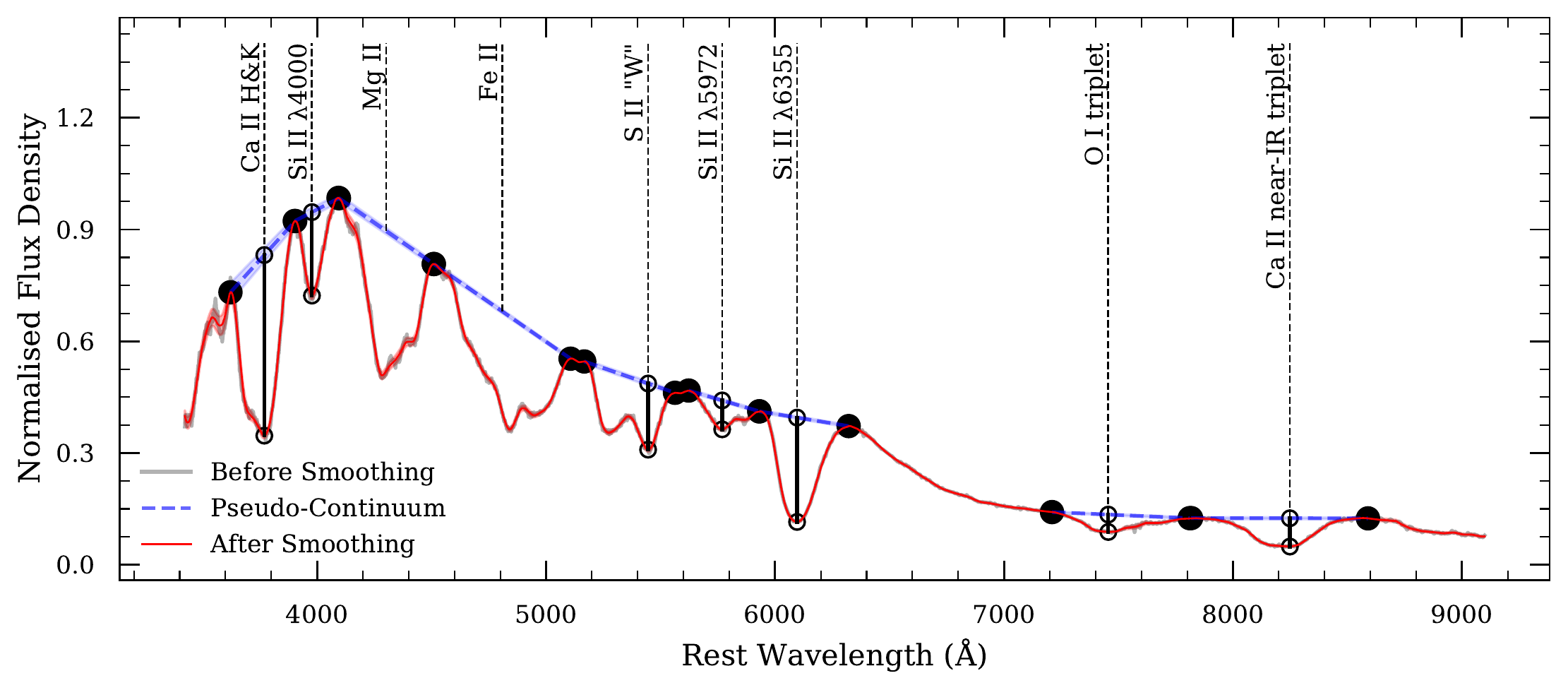}
	\caption{Spectrum of SN 2016coj at roughly 0.5 day before maximum brightness after processing with {\tt respext}. The Galactic reddening-corrected and deredshifted (but otherwise unprocessed) original spectrum appears in grey, while the smoothed spectrum is in red. Large black dots represent the identified feature boundaries which define the pseudo-continuum (in blue). Smaller black circles indicate absorption minima and the pseudo-continuum fluxes at their locations.\label{fig:sample-respext}}
\end{figure*}

Because SN~Ia spectra --- and hence the aforementioned features --- undergo temporal evolution for an individual SN~Ia and exhibit variation over many SNe~Ia (even when comparing similar epochs), the endpoints of each feature must be determined on a spectrum-by-spectrum basis. To this end we have developed \mbox{{\tt respext}}\footnote{\url{https://github.com/benstahl92/respext}}, a {\tt Python} package for automated SN~Ia spectral feature analysis that is an object-oriented and extensively modified refactorisation (or \emph{redux}) of the {\tt spextractor}\footnote{\url{https://github.com/astrobarn/spextractor}} package. Given an input spectrum, the program smooths\footnote{In tests, we have found negligible difference between measurements conducted with and without smoothing. Smoothing does, however, allow us to study spectra whose SNRs would otherwise make their measurement unreliable.} it using a Savitzky-Golay filter \citep{SGfilter} and then automatically (or if necessary, manually) selects absorption-feature boundaries, from which pseudo-continua are derived. It then measures the pEWs, expansion velocities, and boundaries of those features. Figure~\ref{fig:sample-respext} shows the result of this procedure when applied to a spectrum of SN~2016coj near maximum brightness. In the following subsections, we describe our measurement procedure in detail and present our results.

\subsubsection{Pseudo-continua and Pseudo-equivalent Widths}
\label{sssec:pc-and-pew}

After taking steps to standardise\footnote{The steps performed to homogenise input spectra include correcting for Milky Way (MW) reddening using the values given in Table~\ref{tab:SN-information} and assuming the extinction law of \citet{Cardelli1989} as modified by \citet{ext94}, deredshifting (again using values from Table~\ref{tab:SN-information}), flux-normalising, and smoothing.} an input spectrum, the first task is to determine the edges of each of its features. We do this by means of a two-step process: (i) we compute the derivative of the smoothed spectrum and identify the wavelengths corresponding to where it changes from positive to negative (i.e., the wavelengths of local maxima); (ii) of these identified wavelengths, the one corresponding to the maximum flux of the smoothed spectrum within the blue (red) edge boundary (as given in Table~\ref{tab:spectral-features}) is used to define the blue (red) edge of the absorption feature. Owing to the fact that the blue end of the \ion{O}{i} triplet rarely reaches a local maximum, we follow S12b by modifying our procedure to identify where the derivative passes through $-2.0 \times 10^{-18}$ erg s$^{-1}$ cm$^{-2}$~\AA$^{-2}$ (moving in the positive direction). We visually inspect all feature boundaries derived from this procedure, and infrequently override them by manually selecting boundary points when the original ones are not correct. The uncertainty in the flux at the boundary points is assigned as the root-mean-square error (RMSE) between the input and smoothed fluxes within a range identical to the width of the smoothing window centred at the identified boundary wavelengths. We list all measured feature-boundary fluxes (and their uncertainties) in Table~\ref{tab:spec-feature-meas}.

\begin{table*}
\caption{SN~Ia spectral feature measurements near maximum brightness.\label{tab:spec-feature-meas}}
\begin{tabular}{lcrrrrr}
\hline
\hline
\multicolumn{1}{l}{SN} & \multicolumn{1}{c}{Feature} & \multicolumn{1}{r}{$t_{\rm LC}$$^a$} & \multicolumn{1}{c}{$F_\mathrm{b}^b$} & \multicolumn{1}{c}{$F_\mathrm{r}^b$} & \multicolumn{1}{c}{pEW$^{c}$} & \multicolumn{1}{c}{$v^d$}\\
\hline
   SN 2008hv &               Ca II H\&K &   $14.29$ &     $2.66 \pm 0.17$ &     $4.04 \pm 0.11$ &    $62.7 \pm 5.2$ &  $-11.74 \pm 0.16$ \\
    SN 2009D &               Ca II H\&K &   $-5.61$ &     $6.67 \pm 0.74$ &     $7.29 \pm 0.50$ &   $120.1 \pm 6.3$ &  $-20.01 \pm 0.16$ \\
    SN 2009Y &               Ca II H\&K &    $5.72$ &     $8.75 \pm 0.14$ &    $16.24 \pm 0.25$ &   $114.7 \pm 2.4$ &  $-17.87 \pm 0.16$ \\
   SN 2009bv &               Ca II H\&K &   $12.46$ &     $0.86 \pm 0.04$ &     $1.24 \pm 0.04$ &    $83.0 \pm 3.5$ &  $-12.14 \pm 0.16$ \\
   SN 2009cz &               Ca II H\&K &   $-3.03$ &     $6.51 \pm 0.16$ &     $6.75 \pm 0.11$ &   $117.1 \pm 2.7$ &  $-19.59 \pm 0.16$ \\
   SN 2009dc &               Ca II H\&K &   $-5.74$ &     $7.15 \pm 0.11$ &     $6.97 \pm 0.34$ &    $40.4 \pm 4.7$ &  $-16.16 \pm 0.16$ \\
   SN 2009ds &               Ca II H\&K &    $8.62$ &    $12.71 \pm 0.42$ &    $15.66 \pm 0.31$ &    $80.5 \pm 2.9$ &  $-12.61 \pm 0.16$ \\
   SN 2009eu &               Ca II H\&K &   $-4.80$ &     $0.42 \pm 0.02$ &     $0.58 \pm 0.02$ &   $100.5 \pm 3.5$ &  $-18.79 \pm 0.16$ \\
   SN 2009fw &               Ca II H\&K &    $5.30$ &     $0.31 \pm 0.11$ &     $0.43 \pm 0.09$ &   $68.5 \pm 14.5$ &  $-19.88 \pm 0.16$ \\
   SN 2009fw &               Ca II H\&K &    $6.44$ &     $0.53 \pm 0.05$ &     $1.26 \pm 0.05$ &    $77.9 \pm 5.4$ &  $-19.10 \pm 0.16$ \\
\hline
\multicolumn{7}{p{11cm}}{Abridged table of SN~Ia spectral feature measurements (the full table is available as online supplementary material).}\\
\multicolumn{7}{p{11cm}}{$^a$Phases are in rest-frame days as given in Table~\ref{tab:spectra-information}.}\\
\multicolumn{7}{p{11cm}}{$^b$Fluxes at feature boundaries are in units of $10^{-15}$ erg s$^{-1}$ cm$^{-2}$~\AA$^{-1}$.}\\
\multicolumn{7}{p{11cm}}{$^c$Pseudo-equivalent widths are in units of~\AA.}\\
\multicolumn{7}{p{11cm}}{$^d$Expansion velocities are in units of $10^3$ km s$^{-1}$ and are \emph{blueshifts}.}
\end{tabular}
\end{table*}

If the blue and red boundaries of a feature are successfully determined, we define the pseudo-continuum by connecting the boundary points with a line. The lower (upper) uncertainty of the pseudo-continuum is derived by connecting a line between the boundary points, with their fluxes reduced (increased) by their uncertainties. Once the pseudo-continuum is determined, we calculate the pEW \citep[e.g.,][S12b]{Garavini07},
\begin{equation}
	{\rm pEW} = \sum_{i = 0}^{N - 1} \Delta \lambda_i \left(1 - \frac{f(\lambda_i)}{f_c(\lambda_i)}\right),
\end{equation}
where $N$ is the number of pixels between the blue and red boundaries of the feature (which also define the pseudo-continuum as discussed above), $\lambda_i$ ($\Delta \lambda_i$) is the wavelength (width) of the $i$th pixel, and $f(\lambda_i)$ [$f_c(\lambda_i)$] is the spectrum [pseudo-continuum] flux at $\lambda_i$. The uncertainty in our measurement of the pEW is calculated using standard techniques of error propagation using both the uncertainty of the pseudo-continuum (as described above) and the uncertainty in the spectrum flux at each pixel (derived using the RMSE as done for feature boundaries, also described above).

\begin{figure*}
	\centering
	\includegraphics[width=0.9\textwidth]{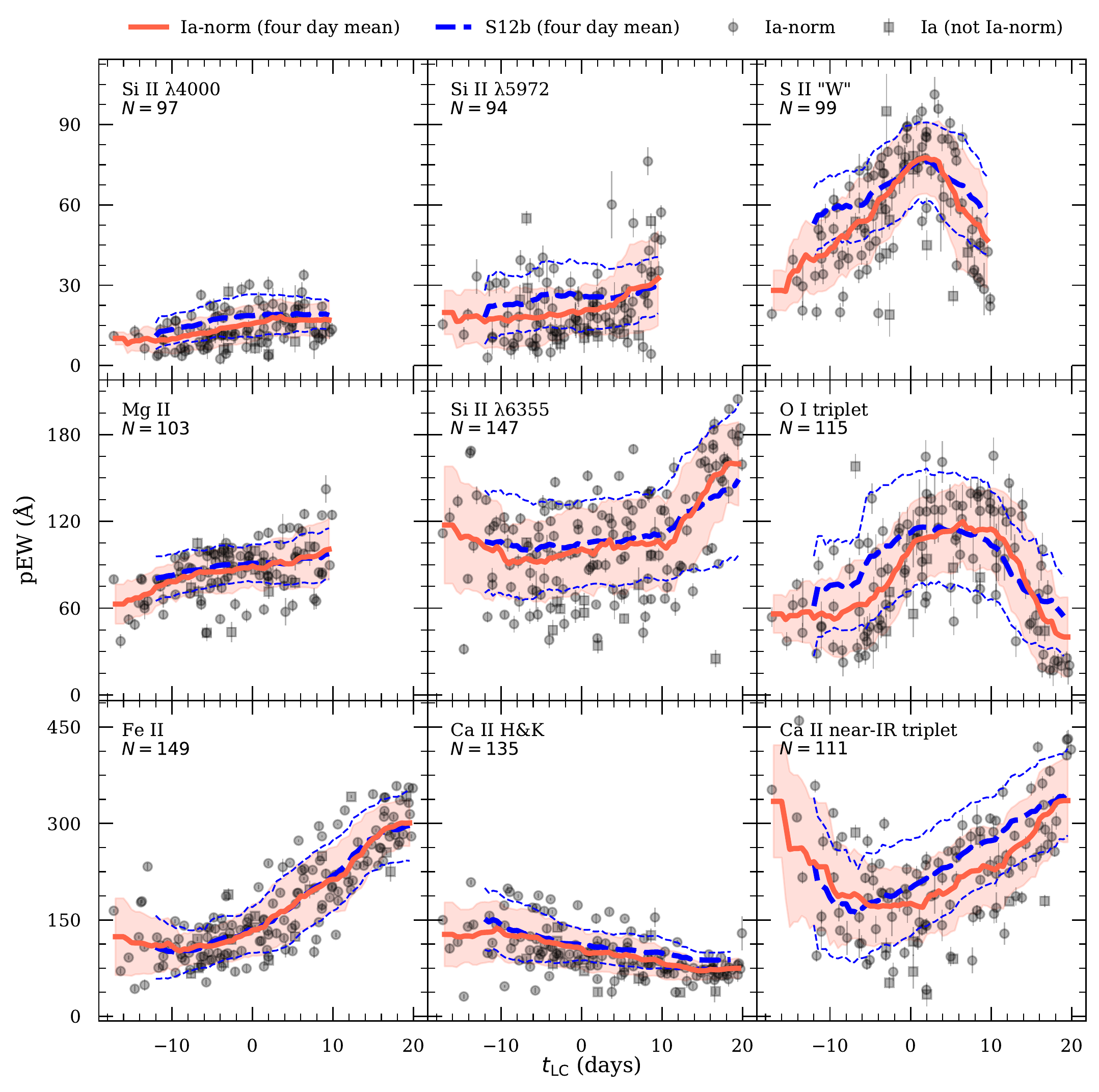}
	\caption{Evolution of pEWs for the features noted in Table~\ref{tab:spectral-features}, grouped by pEW magnitude. Grey circles are measured pEWs for spectra belonging to SNe classified as ``Ia-norm'' according to the prescription of Section~\ref{sec:classification}, and grey squares are those for SNe classified as ``Ia'' of any subtype (except ``Ia-norm'') or ``Ia'' with no subtype determined. The red line and filled region represent the mean and standard deviation (respectively) of all ``Ia-norm'' measurements within four days of each half-day increment in the evolution, and the dashed blue lines represent the corresponding descriptors derived from the dataset of S12b. The feature and number of ``Ia-norm'' measurements for it are included in each panel.\label{fig:pEW}}
\end{figure*}

Table~\ref{tab:spec-feature-meas} includes a column containing all measured pEWs, and we visualise their temporal evolution in Figure~\ref{fig:pEW}. In the same figure, we compare the aggregate pEW evolution for each feature in our dataset to those derived from the dataset of S12b. Given that this comparison is for measurements made between different (but similarly targeted, observed, and reduced) spectra from different SNe~Ia, we find the level of consistency satisfactory. Indeed, the same evolutionary trends clearly manifest themselves in both datasets --- we mention some of the more noteworthy observations in the following paragraphs.

Both the \ion{Ca}{ii} H\&K feature and \ion{Ca}{ii} near-infrared (IR) triplet exhibit relatively large pEWs for $t_{\rm LC}\lesssim -5$ days, but while the former has a pEW that slowly declines through its evolution beyond this point (with noticeably reducing scatter), the pEW of the latter markedly grows. These features, together with the \ion{Fe}{ii} complex (which seems to grow quadratically for $t_{\rm LC} \gtrsim -8$ days), have the largest pEWs of all features measured (and are thus in the last row of Figure~\ref{fig:pEW}).

The \ion{Mg}{ii} feature pEW measurements show a modestly increasing trend and have relatively small scatter compared to those for the \ion{Si}{ii} $\uplambda 6355$ feature and \ion{O}{i} triplet (all three displayed in the central row of Figure~\ref{fig:pEW} owing to their similar range of values). The \ion{O}{i} triplet's mean pEW evolution appears to consist of several distinct stages: there is an increase for $t_{\rm LC} \lesssim 5$ days, at which point the evolution reaches a broad peak of $\sim 120$~\AA, and then there is a stage of decrease. The mean pEW evolution of the \ion{S}{II} ``W'' feature follows a similar trend, except that the peak of $\sim 80$~\AA\ occurs a few days earlier and is more sharply defined.

\ion{Si}{ii} $\uplambda 6355$, the most characteristic spectral feature of SNe~Ia near maximum brightness, shows relatively flat pEW evolution ($\sim 100$~\AA) for $t_{\rm LC} \lesssim 10$ days, after which our measurements are consistent with a ``hint of sharp upturn'' as was noted by S12b, and which is likely due to blending with \ion{Si}{ii} $\uplambda 5972$ at such epochs. Similarly, the \ion{Si}{ii} $\uplambda 5972$ feature exhibits relatively constant (if slightly increasing) pEW evolution until $t_{\rm LC} \approx 5$ days, at which phase there is an uptick, likely due to blending with the \ion{Na}{i}~D line (from the MW, and owing to their low redshifts, possibly from the host galaxies of the SNe). The \ion{Si}{ii} $\uplambda 4000$ feature has the lowest aggregate pEWs in our sample (hence its position in the first row of Figure~\ref{fig:pEW}, along with the measurements for \ion{Si}{ii} $\uplambda 5972$ and \ion{S}{ii} ``W''), and shows evidence for a slight trend of increasing pEW.

\subsubsection{Expansion Velocities}
\label{sssec:exp-velocity}

With feature boundaries determined according to Section~\ref{sssec:pc-and-pew}, we identify the absorption minimum (wavelength and flux) in each feature by fitting the smoothed flux (within each feature boundary) with a cubic spline and computing the minimum. We do this for all features with identified boundaries except for \ion{Mg}{ii} and \ion{Fe}{ii}, which are composites of so many blended lines that there is ambiguity when choosing a reference wavelength against which to measure expansion velocities. The \ion{S}{ii} ``W'' feature has two broad absorptions, so for consistency --- both internally and with the results described by S12b --- we always measure the minimum of the redder of the two features (even if the bluer component has a deeper absorption). As with measurements of feature boundaries, we perform a visual inspection, and in cases where the spline fit does not accurately reflect the true flux minimum we manually adjust the range over which the spline is fit in order to more faithfully capture the signal. Following S12b, we impose a 2~\AA\ uncertainty on the wavelength of the feature minimum (and do not explicitly account for systematic uncertainties due to the spectral resolution).

\begin{figure*}
	\centering
	\includegraphics[width=0.9\textwidth]{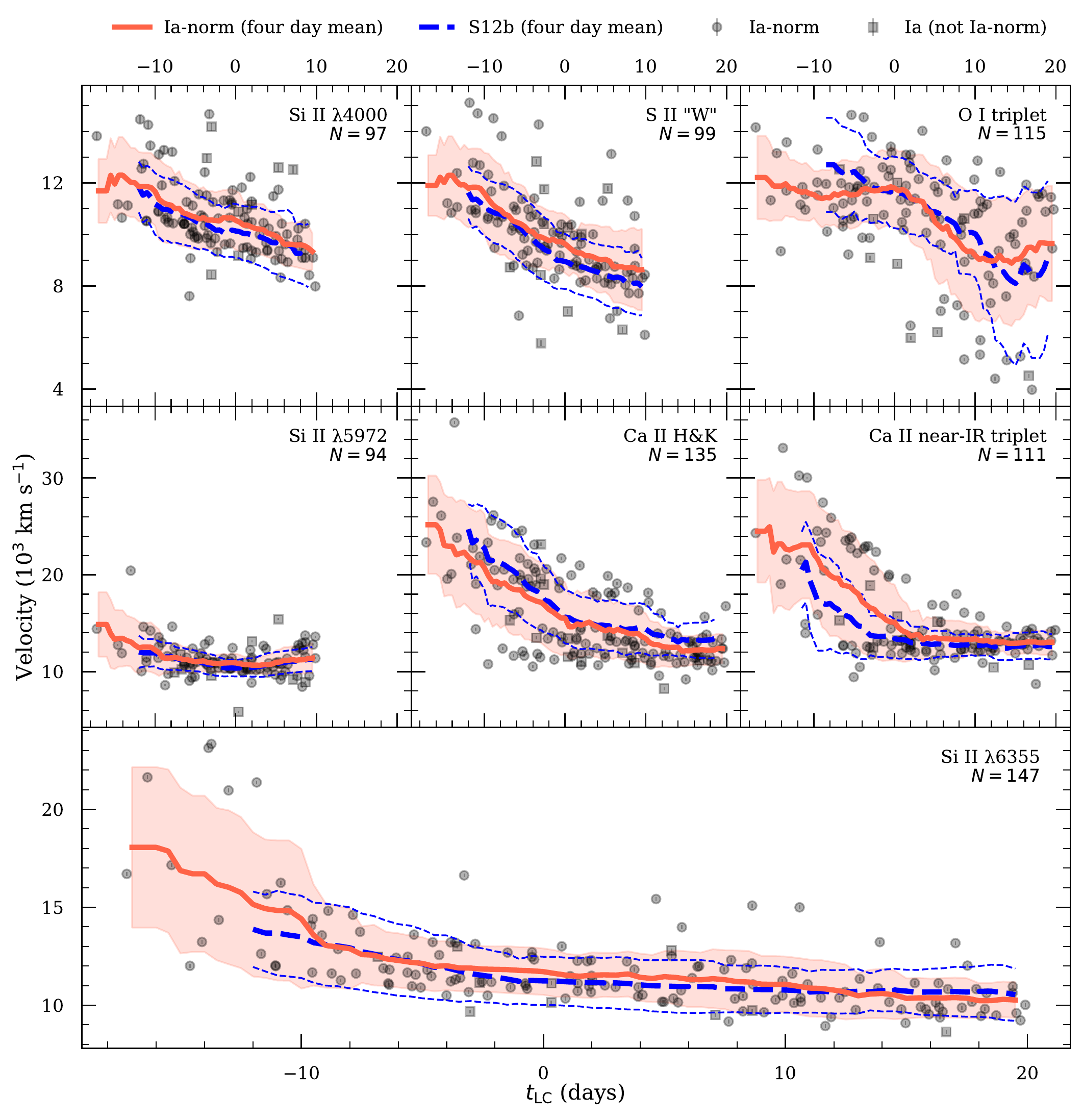}
	\caption{Same as Figure~\ref{fig:pEW}, but for expansion velocities. Though we show positive velocity scales, we reiterate that the velocities correspond to \emph{blueshifts}.\label{fig:vexp}}
\end{figure*}

To calculate the expansion velocity of a feature, $v$, we use the wavelength of its flux minimum (as determined above) and the appropriate rest wavelength (as given in Table~\ref{tab:spectral-features}) with the relativistic Doppler equation. The uncertainty in the expansion velocity is obtained by propagating the wavelength uncertainty (as described above). We present all of our velocity measurements in Table~\ref{tab:spec-feature-meas}. We emphasise that they are derived from \emph{blueshifted} spectral features (and hence appear as negative number in the table). All velocity measurements are shown in Figure~\ref{fig:vexp}, and the aggregate results are compared to those derived from the dataset of S12b. As with the pEW comparison, we find clear qualitative consistency in evolutionary trends (especially given some allowance for biases due to low-number statistics at the earliest epochs).

Similar to S12b, we find the highest expansion velocities from the two \ion{Ca}{ii} features we investigated. The features exhibit similar evolution, with velocities in excess of $25,000$ km s$^{-1}$ (and as high as $\sim 30,000$ km s$^{-1}$) for $t \lesssim -5$ days, followed by a rapid decline to relative constancy (slightly decreasing for \ion{Ca}{ii} H\&K) at $\sim 12,000$ km s$^{-1}$ for $t \gtrsim 0$ days.

All three features of \ion{Si}{ii} show a similar evolutionary track of modest decline, albeit with different scales. The largest velocities are claimed by \ion{Si}{ii} $\uplambda 6355$, followed by \ion{Si}{ii} $\uplambda 5972$ (both of which converge to a steady velocity of $\sim 11,000$ km s$^{-1}$ for $t \gtrsim 0$ days), and finally \ion{Si}{ii} $\uplambda 4000$ (which continues to decline throughout the evolution). The velocity of the \ion{S}{ii} ``W'' feature shows a very similar evolution to that of \ion{Si}{ii} $\uplambda 4000$, but with a bit more scatter and a slightly steeper decline.

The expansion velocities of the \ion{O}{i} triplet cover a similar range of values to those of the \ion{Si}{ii} ``W'' feature, but with a significantly larger degree of scatter (especially for later epochs). This is unsurprising: the \ion{O}{i} triplet is a broad feature, and thus when it becomes weak (as it does at later phases, as shown in Figure~\ref{fig:pEW}) the exact location of the minimum is more challenging to robustly determine. It is difficult to quantify the extent to which this mechanism introduces scatter relative to what may be intrinsic, but after visually inspecting the results, we find the derived minima to be reasonable.

\subsection{Late-time Spectra}

As a SN~Ia reaches the so-called nebular phase in its evolution (starting $t \gtrsim 100$ days after maximum light, and fully for $t \gtrsim 160$ days), the density of the ejecta diminishes to the point of becoming optically thin, thereby allowing light from deep within the interior to escape. This results in broad emission lines (due mostly to iron-group elements) in the late-time optical spectra of SNe~Ia, which may encode important physical and geometric details of the explosion mechanism(s) \citep{Maeda2010b,Maguire2018}. In particular, many studies of late-time SN~Ia spectra have considered three broad emission features centred near 4701, 6155, and 7378~\AA, which are attributed to blends of various lines of [\ion{Fe}{iii}], [\ion{Fe}{ii}], and [\ion{Ni}{ii}], respectively \citep{Mazzali1998,Maeda2010a,Blondin2012,bsnipV,Maguire2018}.

Of the spectra in our dataset having light-curve-determined phases, there are 15 (spanning 7 SNe~Ia) for which $t_\mathrm{LC} \geq 160$ days. Though this is not a sufficiently large sample to perform a stand-alone study (and because a subset of these spectra have already been considered in other works; see the references listed in Table~\ref{tab:spectra-information}), we perform only a brief analysis focusing on the velocity shift of the [\ion{Fe}{iii}] $\uplambda 4701$ feature and the mean velocity shift of the two remaining features (which, for consistency with the aforementioned studies, we refer to as the ``nebular velocity''). We describe our methodology and measurements in the following subsections.

\subsubsection{Methodology}

We measure velocities of the listed features in our nebular spectra using tools from our {\tt respext} package. Again, we preprocess spectra by correcting for Galactic extinction and then deredshifting, flux-normalising, and smoothing. Emission peaks are identified by eye and then, following the approach described in Section~\ref{sssec:exp-velocity}, we fit a cubic spline to the smoothed spectrum in the vicinity of the peak, allowing us to derive the wavelength at which the flux is maximal. Consistent with our treatment of early-time spectra, we impose a uniform 2~\AA\ uncertainty on all wavelengths determined by this method. The velocity of the feature is then obtained using the relativistic Doppler equation. Our results are summarised in Table~\ref{tab:spec-feature-meas-late}.
\begin{table}
\caption{Late-time SN~Ia spectral feature measurements.\label{tab:spec-feature-meas-late}}
\setlength{\tabcolsep}{0.5em}
\begin{tabular}{lcrrr}
\hline
\hline
\multicolumn{1}{l}{SN} & \multicolumn{1}{c}{$t_{\rm LC}$$^a$} & \multicolumn{1}{c}{Velocity$^b$} & \multicolumn{1}{c}{Velocity$^b$} & \multicolumn{1}{c}{Velocity$^b$}\\
\multicolumn{1}{l}{Name} & \multicolumn{1}{r}{} & \multicolumn{1}{c}{[\ion{Fe}{iii}] $\uplambda 4701$} & \multicolumn{1}{c}{[\ion{Fe}{ii}] $\uplambda 7155$} & \multicolumn{1}{c}{[\ion{Ni}{ii}] $\uplambda 7378$}\\
\hline
   SN 2009ig &  160.3 &  $-2.47 \pm 0.13$ &   $0.83 \pm 0.08$ &  $-1.36 \pm 0.08$ \\
   SN 2011by &  206.7 &  $-1.68 \pm 0.13$ &  $-1.38 \pm 0.08$ &  $-1.73 \pm 0.08$ \\
   SN 2011by &  310.3 &  $-0.54 \pm 0.13$ &  $-1.05 \pm 0.08$ &  $-1.15 \pm 0.08$ \\
   SN 2011fe &  164.9 &  $-2.64 \pm 0.13$ &  $-1.37 \pm 0.08$ &               ... \\
   SN 2011fe &  203.9 &  $-1.54 \pm 0.13$ &  $-1.26 \pm 0.08$ &  $-1.12 \pm 0.08$ \\
   SN 2011fe &  224.8 &  $-1.17 \pm 0.13$ &  $-0.93 \pm 0.08$ &  $-1.09 \pm 0.08$ \\
   SN 2011fe &  309.6 &  $-0.63 \pm 0.13$ &  $-1.22 \pm 0.08$ &  $-0.94 \pm 0.08$ \\
   SN 2011fe &  346.5 &  $-0.32 \pm 0.13$ &  $-0.76 \pm 0.08$ &  $-0.89 \pm 0.08$ \\
   SN 2011fe &  378.5 &   $0.20 \pm 0.13$ &  $-1.10 \pm 0.08$ &  $-0.90 \pm 0.08$ \\
   SN 2013dy &  422.2 &   $2.88 \pm 0.13$ &   $0.55 \pm 0.08$ &   $0.39 \pm 0.08$ \\
   SN 2013gy &  271.8 &  $-0.56 \pm 0.13$ &  $-0.29 \pm 0.08$ &  $-2.95 \pm 0.08$ \\
    SN 2014J &  264.6 &  $-0.90 \pm 0.13$ &   $0.86 \pm 0.08$ &   $1.45 \pm 0.08$ \\
    SN 2014J &  291.6 &  $-0.75 \pm 0.13$ &   $1.00 \pm 0.08$ &   $1.38 \pm 0.08$ \\
 ASASSN 14lp &  170.1 &  $-2.07 \pm 0.13$ &   $0.57 \pm 0.08$ &   $0.53 \pm 0.08$ \\
 ASASSN 14lp &  175.1 &  $-2.23 \pm 0.13$ &   $0.44 \pm 0.08$ &   $0.19 \pm 0.08$ \\
\hline
\multicolumn{5}{p{8cm}}{$^a$Spectral phases are in rest-frame days as given in Table~\ref{tab:spectra-information}.}\\
\multicolumn{5}{p{8cm}}{$^b$Velocities are in units of $10^3$ km s$^{-1}$. Negative values are blueshifted. Systematic uncertainties associated with the resolution of the spectra are not included.}
\end{tabular}
\end{table}

\subsubsection{[\ion{Fe}{iii}] $\uplambda 4701$ Velocities}
\label{sssec:feiii-velocity}

\begin{figure}
	\centering
	\includegraphics[width=\columnwidth]{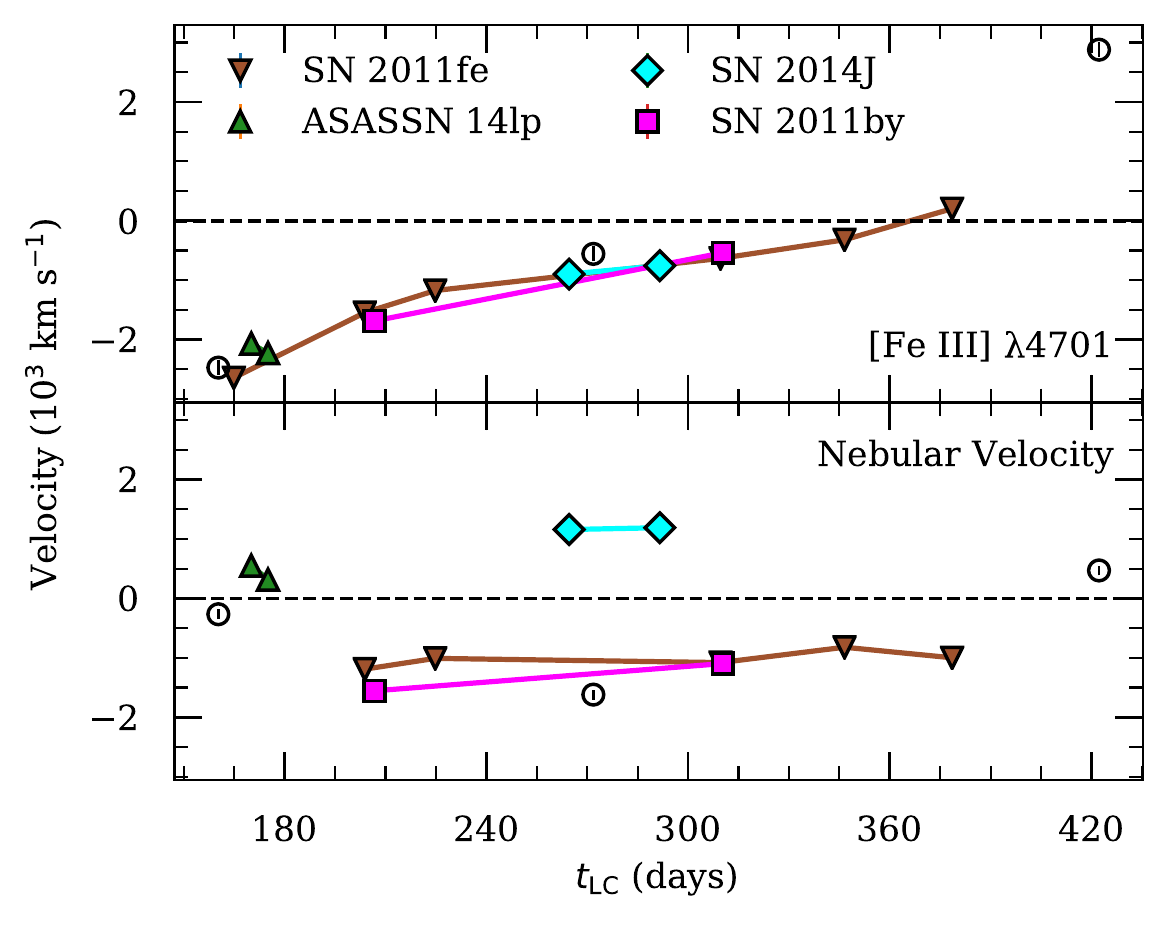}
	\caption{Measured velocity shifts from the nebular spectra in our sample. The top panel shows the velocity shifts for the [\ion{Fe}{iii}] $\uplambda 4701$ feature, while the bottom shows the nebular velocities (as discussed in Section~\ref{sssec:nebular-velocity}). SNe~Ia with multiple nebular spectra are marked as indicated in the legend and connected by lines. The error bars, which do not account for systematic uncertainties from the resolution of our spectra, are typically smaller than the markers. For a typical resolution of $\sim 10$~\AA, the omitted systematic uncertainty amounts to $\sim 500$ km s$^{-1}$.\label{fig:late-time-velocities}}
\end{figure}

We present our measurements of the velocity shifts of the [\ion{Fe}{iii}] $\uplambda 4701$ feature in the top panel of Figure~\ref{fig:late-time-velocities}. Similar to \citet{bsnipV}, we find evidence for a slow decrease in blueshift (i.e., a velocity \emph{increase}) in the nebular-phase evolution. For the three SNe~Ia in our sample having multiple nebular-phase spectra with nonnegligible temporal separation (SN~2011fe, SN~2011by, and SN~2014J), we find average velocity increase rates of 15, 11, and 5~km s$^{-1}$ d$^{-1}$ (respectively).

\subsubsection{Nebular Velocities}
\label{sssec:nebular-velocity}

As with \citet{Maeda2010a}, \citet{Blondin2012}, and \citet{bsnipV}, we derive nebular velocities as the arithmetic mean of the [\ion{Fe}{ii}] $\uplambda 7155$ and [\ion{Ni}{ii}] $\uplambda 7378$ feature velocities. Whereas previous studies have determined the uncertainty in the nebular velocity as the difference between the constituent velocities \citep{Maeda2010a}, or half of this difference \citep{bsnipV}, we derive it from direct propagation of uncertainties. We present our nebular velocity measurements in the bottom panel of Figure~\ref{fig:late-time-velocities}. In contrast to the slow (but noticeable) increasing trend in the [\ion{Fe}{iii}] $\uplambda 4701$ velocities, we find an even weaker trend in nebular velocities. For the previously mentioned set of three SNe~Ia with multiple nebular spectra, we find average velocity increase rates of just 2, 4, and 1~km s$^{-1}$ d$^{-1}$ --- consistent with the assertion made by \citet{bsnipV} that a single measurement of the nebular velocity of a given SN~Ia is sufficient to describe that SN throughout its nebular-phase evolution.

\section{Conclusion}
\label{sec:conclusion}

In this paper we present 637 optical spectra collected by the Berkeley Supernova Ia Program using the Kast double spectrograph at Lick Observatory and LRIS at the W. M. Keck Observatory between 2009 and 2018. Careful observation and processing techniques perfected over the last 20+ years are employed to prepare the spectra in a manner that is (i) self-consistent and (ii) consistent with earlier BSNIP spectral data releases (S12a).

We employ a robust automated spectral classification procedure that uses {\tt SNID} to derive the type, subtype, redshift, and rest-frame phase of the spectra in our dataset, achieving a successful result in the majority of cases. Furthermore, we perform a study of the results and conclude that failures preferentially occur for late-phase spectra (where the temporal coverage of {\tt SNID} is sparse) and for spectra with lower SNRs (and which are thus of lower quality). Where independent measurements (i.e., host-galaxy redshifts, and light-curve-derived rest-frame phases) are available, we compare them to {\tt SNID}-based predictions. The redshifts show negligible difference in aggregate and have relatively small scatter, while the phases have a larger --- but still reasonable --- scatter (especially when a more temporally restrictive subset is selected). After combining the classifications in cases where multiple spectra are available for a given object, we address the several cases in which a selected object was not classified as a SN~Ia. Ultimately, we obtain a final sample of 626 spectra from 242 low-redshift SNe~Ia.

We study the early-time and late-time properties of our dataset, with emphasis on measurements of the most prominent features in SN~Ia spectra at such phases. In particular, we measure the expansion velocities, pEWs, and fluxes at the boundaries of nine absorption-feature complexes from the subset of our spectra that were observed within 20 days of maximum light. When we compare with the analogous set of measurements performed on an earlier set of BSNIP spectra (S12b), we find clear evidence for the same evolutionary behaviours in the features. Similarly, we measure the velocity shifts of three emission features from the subset of our spectra that were observed more than 160 days after maximum light. With just 15 such nebular spectra, our sample is too small to merit a stand-alone study, but we do find clear manifestations of the evolutionary behaviours noted by more comprehensive studies.

When our dataset is combined with that described by S12a, the BSNIP low-redshift SN~Ia spectral dataset reaches nearly 2000 optical spectra, all of which have been handled consistently through all phases of observing and processing. Further utility will be unlocked by considering the aforementioned spectral dataset in conjunction with its companion photometric dataset of more than 250 SNe~Ia from the Lick Observatory Supernova Search follow-up program (see G10 and S19, for the photometric datasets covering 1998--2008 and 2009--2018, respectively). In a future study, we will leverage these datasets to explore the extent to which photometrically derived parameters can be reconstructed from SN~Ia spectra (Stahl et al., in prep.).

\section*{Acknowledgements}

We thank
Bela Abolfathi,
Louis Abramson,
Iair Arcavi,
Roberto Assef,
Aaron Barth,
Vardha Bennert,
Andrew Bigley,
Peter Blanchard,
Joshua Bloom,
Benjamin Boizelle,
Azalee Bostroem,
Andrew Brandel,
Michael Busch,
Zheng Cai,
Gabriela Canalizo,
Dan Carson,
Jieun Choi,
Daniel Cohen,
Michael Cooper,
Maren Cosens,
Antonino Cucchiara,
Aleks Diamond-Stanic,
Subo Dong,
Sean Fillingham,
Ryan Foley,
Mohan Ganeshalingam,
Elinor Gates,
Jenny Greene,
Christopher Griffith,
Kyle Hiner,
Sebastian Hoenig,
Griffin Hosseinzadeh,
Yiseul Jeon,
Caitlin Johnson,
Daniel Kasen,
Minkyu Kim,
Mariana Lazarova,
David Levitan,
Matthew Malkan,
Christina Manzano-King,
Bruce Margon,
Carl Melis,
Allison Merritt,
Adam Miller,
Maryam Modjaz,
Adam Morgan,
Alekzandir Morton,
Robin Mostardi,
My Nguyen,
Peter Nugent,
Liuyi Pei,
Daniel Perley,
Dovi Poznanski,
Armin Rest,
Jacob Rex,
Roger Romani,
Liming Rui,
Frank Serduke,
Remington Sexton,
Jaejin Shin,
Marijana Smailagic,
Alessandro Sonnenfeld,
Thea Steele,
Tommaso Treu,
Vivian U,
Stefano Valenti,
Alex Vogler,
Jonelle Walsh,
Marie Wingyee Lau,
Gabor Worseck,
Fang Yuan,
Sameen Yunus,
Yinan Zhu,
and others for their assistance with some of the
observations over the past decade presented in this paper.
We would also like to express our gratitude to the staffs at the Lick
and Keck Observatories for their support, and our anonymous referee whose careful reading and constructive comments improved the manuscript.  KAIT and its ongoing
operation were made possible by donations from Sun Microsystems, Inc.,
the Hewlett-Packard Company, AutoScope Corporation, Lick Observatory,
the NSF, the University of California, the Sylvia \& Jim Katzman
Foundation, and the TABASGO Foundation.

A major upgrade of the Kast spectrograph on the Shane 3~m telescope
at Lick Observatory was made possible through generous gifts from
William and Marina Kast as well as the Heising-Simons Foundation.
Research at Lick Observatory is partially supported by a generous gift from
Google. Support for A.V.F.'s supernova group has also been provided by the
NSF, Marc J. Staley (B.E.S. is a Marc J. Staley Graduate Fellow), the
Richard and Rhoda Goldman Fund, the TABASGO Foundation, Gary and Cynthia
Bengier (T.deJ. is a Bengier Postdoctoral Fellow), the Christopher R.
Redlich Fund, and the Miller Institute for Basic Research in Science
(U.C. Berkeley). In addition, we greatly appreciate contributions from
numerous individuals, including
Charles Baxter and Jinee Tao,
George and Sharon Bensch,
Firmin Berta,
Marc and Cristina Bensadoun,
Frank and Roberta Bliss,
Eliza Brown and Hal Candee,
Kathy Burck and Gilbert Montoya,
Alan and Jane Chew,
David and Linda Cornfield,
Michael Danylchuk,
Jim and Hildy DeFrisco,
William and Phyllis Draper,
Luke Ellis and Laura Sawczuk,
Jim Erbs and Shan Atkins,
Alan Eustace and Kathy Kwan,
Peter and Robin Frazier,
David Friedberg,
Harvey Glasser,
John and Stacey Gnuse,
Charles and Gretchen Gooding,
Alan Gould and Diane Tokugawa,
Thomas and Dana Grogan,
Timothy and Judi Hachman,
Alan and Gladys Hoefer,
Charles and Patricia Hunt,
Stephen and Catherine Imbler,
Adam and Rita Kablanian,
Loren and Kristen Kinczel,
Roger and Jody Lawler,
Kenneth and Gloria Levy,
Peter Maier,
DuBose and Nancy Montgomery,
Rand Morimoto and Ana Henderson,
Sunil Nagaraj and Mary Katherine Stimmler,
Peter and Kristan Norvig,
James and Marie O'Brient,
Emilie and Doug Ogden,
Paul and Sandra Otellini,
Jeanne and Sanford Robertson,
Paul Robinson,
Sissy Sailors and Red Conger,
Stanley and Miriam Schiffman,
Thomas and Alison Schneider,
Ajay Shah and Lata Krishnan,
Alex and Irina Shubat,
the Silicon Valley Community Foundation,
Bruce and Deborah Smith,
Mary-Lou Smulders and Nicholas Hodson,
Hans Spiller,
Alan and Janet Stanford,
the Hugh Stuart Center Charitable Trust,
Clark and Sharon Winslow,
Weldon and Ruth Wood,
David and Angie Yancey, 
Thomas Zdeblick 
and many others.

B.E.S. thanks Marc J. Staley for generously providing fellowship funding, and S. L. Watkins and C. W. Fink for their helpful suggestions during the development of {\tt respext}. M.L.G. acknowledges support from the DIRAC Institute in the Department of Astronomy at the University of Washington. The DIRAC Institute is supported through generous gifts from the Charles and Lisa Simonyi Fund for Arts and Sciences, and the Washington Research Foundation. X.G.W. is supported by the National Natural Science Foundation of China (NSFC grant 11673006), and the Guangxi Science Foundation (grants 2016GXNSFFA380006 and 2017AD22006). X.W. is supported by the National Natural Science Foundation of China (NSFC grants 11325313, 11633002, and 11761141001), and the National Program on Key Research and Development Project (grant 2016YFA0400803).

This research has made use of the NASA/IPAC Extragalactic Database (NED), which is operated by the Jet Propulsion Laboratory, California Institute of Technology, under contract with NASA. Some of the data presented herein were obtained at the W. M. Keck Observatory, which is operated as a scientific partnership among the California Institute of Technology, the University of California, and the National Aeronautics and Space Administration (NASA). The Observatory was made possible by the generous financial support of the W. M. Keck Foundation. The authors wish to recognise and acknowledge the very significant cultural role and reverence that the summit of Maunakea has always had within the indigenous Hawaiian community. We are most fortunate to have the opportunity to conduct observations from this mountain.




\bibliographystyle{mnras}
\bibliography{Ia_update_bib}




\appendix

\section{Sample Information}
\label{app:sample-info}

\onecolumn
{\footnotesize
\begin{longtable}{lcrrlcccrrc}
\caption{SN~Ia information.\label{tab:SN-information}}\\
\hline
\hline
\multicolumn{1}{l}{SN} & \multicolumn{1}{c}{Discovery} & \multicolumn{1}{r}{R.A.} & \multicolumn{1}{r}{Decl.} & \multicolumn{1}{c}{$z_{\rm helio}$$^a$} & \multicolumn{1}{c}{$E(B-V)_{\rm MW}$$^b$} & \multicolumn{1}{c}{{\tt SNID}$^c$} & \multicolumn{1}{c}{\# of} & \multicolumn{1}{c}{First$^d$} & \multicolumn{1}{c}{Last$^d$} & \multicolumn{1}{c}{MJD$_{\rm max}^e$}\\
\multicolumn{1}{l}{Name} & \multicolumn{1}{c}{Date (UT)} & \multicolumn{1}{r}{$\alpha(2000)$} & \multicolumn{1}{r}{$\delta(2000)$} & \multicolumn{1}{c}{} & \multicolumn{1}{c}{(mag)} & \multicolumn{1}{c}{(sub)type} & \multicolumn{1}{c}{Spectra} & \multicolumn{1}{c}{Epoch} & \multicolumn{1}{c}{Epoch} & \multicolumn{1}{c}{Reference}\\
\hline
\endfirsthead
\hline
\hline
\multicolumn{1}{l}{SN} & \multicolumn{1}{c}{Discovery} & \multicolumn{1}{r}{R.A.} & \multicolumn{1}{r}{Decl.} & \multicolumn{1}{c}{$z_{\rm helio}$$^a$} & \multicolumn{1}{c}{$E(B-V)_{\rm MW}$$^b$} & \multicolumn{1}{c}{{\tt SNID}$^c$} & \multicolumn{1}{c}{\# of} & \multicolumn{1}{c}{First$^d$} & \multicolumn{1}{c}{Last$^d$} & \multicolumn{1}{c}{MJD$_{\rm max}^e$}\\
\multicolumn{1}{l}{Name} & \multicolumn{1}{c}{Date (UT)} & \multicolumn{1}{r}{$\alpha(2000)$} & \multicolumn{1}{r}{$\delta(2000)$} & \multicolumn{1}{c}{} & \multicolumn{1}{c}{(mag)} & \multicolumn{1}{c}{(sub)type} & \multicolumn{1}{c}{Spectra} & \multicolumn{1}{c}{Epoch} & \multicolumn{1}{c}{Epoch} & \multicolumn{1}{c}{Reference}\\
\hline
\endhead
\hline
\multicolumn{11}{c}{Table \thetable~continued}
\endfoot
\hline
\multicolumn{11}{p{18cm}}{$^a$Host-galaxy heliocentric redshifts are from NED unless marked with a ``$\dagger$'' symbol, in which case they are collected using the Open Supernova Catalog \citep{openSNe} from the following sources: (c) the supernova catalog of \citet{2012cat}, (p) the spectroscopic host-galaxy observations of PTF SNe~Ia described by \citet{PTF_host_spec}, (t) the TNS, (w) the Weizmann Interactive Supernova Data Repository \citet[WISeREP][]{WISeREP}, or (d) the appropriate discovery or classification announcement (e.g., CBET or IAUC).}\\
\multicolumn{11}{p{18cm}}{$^b$Extinction is calculated at the SN position using the dust maps of \citet{Schlegel1998} subject to the recalibration of \citet{Schlafly2011}.}\\
\multicolumn{11}{p{18cm}}{$^c$SN classifications are derived from our {\tt SNID} classification scheme, as described in Section~\ref{sec:classification}. Based on the arguments made by \citet{Iax}, we have relabeled all SNe~Ia with a ``Ia-02cx'' subtype as ``Iax''.}\\
\multicolumn{11}{p{18cm}}{$^d$First and last observation epochs are in rest-frame days relative to the time of \emph{B}-band maximum brightness, and are computed using information from the table.}\\
\multicolumn{11}{p{18cm}}{$^e$References for the light-curve-determined MJD corresponding to \emph{B}-band maximum brightness are as follows: (1) S19, (2) \citet{CfAIR} and references therein, (3) \citet{Foundation}, (4) \citet{CSP3}, (5) \citet{Maguire14}, (6) \citet{Khan2011}, (7) \citet{Zhang2014}, (8) \citet{Yamanaka2014}, (9) \citet{Shappee2016}, and (10) \citet{Srivastav2017}.}\\
\endlastfoot
               SN 2008hm &  2008$-$11$-$25 &   $51.7954$ &   $46.9443$ &                        $0.0197$ &  $0.381$ &       Ia &   1 &   $26.5$ &      ... &    2 \\
               SN 2008hv &  2008$-$12$-$02 &  $136.8920$ &    $3.3923$ &                        $0.0125$ &  $0.027$ &  Ia-norm &   1 &   $14.3$ &      ... &    2 \\
               SN 2008hy &  2008$-$12$-$06 &   $56.2852$ &   $76.6654$ &                        $0.0085$ &  $0.203$ &       Ia &   1 &   $32.9$ &      ... &    2 \\
                SN 2009D &  2009$-$01$-$02 &   $58.5951$ &  $-19.1817$ &                        $0.0250$ &  $0.046$ &  Ia-norm &   1 &   $-5.6$ &      ... &    1 \\
                SN 2009Y &  2009$-$02$-$01 &  $220.5990$ &  $-17.2468$ &                        $0.0094$ &  $0.087$ &  Ia-norm &   3 &    $5.7$ &   $62.9$ &    2 \\
                SN 2009V &  2009$-$02$-$02 &  $153.2030$ &   $43.1832$ &  $0.0930$$^{\dagger ({\rm c})}$ &  $0.012$ &  Ia-norm &   1 &      ... &      ... &  ... \\
               SN 2009ae &  2009$-$02$-$15 &  $249.8700$ &   $21.3154$ &                        $0.0311$ &  $0.049$ &  Ia-norm &   1 &      ... &      ... &  ... \\
               SN 2009an &  2009$-$02$-$27 &  $185.6980$ &   $65.8512$ &                        $0.0092$ &  $0.016$ &  Ia-norm &   2 &   $21.1$ &   $40.7$ &    2 \\
               SN 2009bp &  2009$-$03$-$17 &  $211.9570$ &   $36.6436$ &                             ... &  $0.004$ &       Ia &   1 &      ... &      ... &  ... \\
               SN 2009bs &  2009$-$03$-$21 &  $201.7340$ &   $52.7549$ &                        $0.0298$ &  $0.011$ &  Ia-91bg &   2 &      ... &      ... &  ... \\
               SN 2009bv &  2009$-$03$-$27 &  $196.8350$ &   $35.7844$ &                        $0.0367$ &  $0.008$ &  Ia-norm &   1 &   $12.5$ &      ... &    2 \\
               SN 2009cz &  2009$-$04$-$06 &  $138.7500$ &   $29.7353$ &                        $0.0211$ &  $0.022$ &       Ia &   1 &   $-3.0$ &      ... &    4 \\
               SN 2009dc &  2009$-$04$-$09 &  $237.8010$ &   $25.7078$ &                        $0.0214$ &  $0.060$ &  Ia-norm &   8 &   $-5.7$ &  $109.5$ &    1 \\
               SN 2009do &  2009$-$04$-$22 &  $188.7430$ &   $50.8512$ &                        $0.0397$ &  $0.013$ &  Ia-norm &   1 &   $23.4$ &      ... &    2 \\
               SN 2009ds &  2009$-$04$-$28 &  $177.2670$ &   $-9.7291$ &                        $0.0193$ &  $0.033$ &  Ia-norm &   1 &    $8.6$ &      ... &    2 \\
               SN 2009en &  2009$-$05$-$08 &  $221.5940$ &   $13.0242$ &                        $0.0467$ &  $0.020$ &  Ia-norm &   2 &      ... &      ... &  ... \\
               SN 2009ep &  2009$-$05$-$11 &  $208.0440$ &    $2.3242$ &                        $0.0237$ &  $0.025$ &  Ia-norm &   2 &      ... &      ... &  ... \\
               SN 2009eq &  2009$-$05$-$11 &  $280.0350$ &   $40.1268$ &                        $0.0236$ &  $0.053$ &      Ia* &   3 &      ... &      ... &  ... \\
               SN 2009ew &  2009$-$05$-$16 &  $249.7490$ &   $17.9828$ &                             ... &  $0.062$ &  Ia-norm &   2 &      ... &      ... &  ... \\
               SN 2009eu &  2009$-$05$-$21 &  $247.1710$ &   $39.5535$ &                        $0.0304$ &  $0.010$ &  Ia-norm &   1 &   $-4.8$ &      ... &    1 \\
               SN 2009ft &  2009$-$05$-$23 &  $216.0250$ &    $7.7695$ &                        $0.0568$ &  $0.021$ &  Ia-norm &   1 &      ... &      ... &  ... \\
               SN 2009fx &  2009$-$05$-$29 &  $253.2970$ &   $23.9653$ &                        $0.0477$ &  $0.049$ &  Ia-norm &   1 &      ... &      ... &  ... \\
               SN 2009fl &  2009$-$05$-$30 &  $246.2920$ &   $40.8891$ &                        $0.0294$ &  $0.007$ &  Ia-norm &   2 &      ... &      ... &  ... \\
               SN 2009fu &  2009$-$06$-$01 &   $33.0375$ &   $44.5653$ &                        $0.0171$ &  $0.076$ &  Ia-norm &   1 &      ... &      ... &  ... \\
               SN 2009fy &  2009$-$06$-$01 &  $351.0210$ &   $16.6641$ &                        $0.0410$ &  $0.028$ &  Ia-norm &   2 &      ... &      ... &  ... \\
               SN 2009fv &  2009$-$06$-$02 &  $247.4340$ &   $40.8116$ &                        $0.0293$ &  $0.005$ &  Ia-norm &   3 &    $3.8$ &   $16.4$ &    1 \\
               SN 2009gq &  2009$-$06$-$02 &  $333.7200$ &   $17.5131$ &  $0.0670$$^{\dagger ({\rm c})}$ &  $0.037$ &  Ia-norm &   1 &      ... &      ... &  ... \\
               SN 2009fw &  2009$-$06$-$06 &  $308.0770$ &  $-19.7332$ &                        $0.0282$ &  $0.050$ &  Ia-norm &   4 &    $5.3$ &   $17.9$ &    2 \\
               SN 2009gf &  2009$-$06$-$15 &  $213.9050$ &   $14.2802$ &                        $0.0185$ &  $0.022$ &  Ia-norm &   3 &      ... &      ... &  ... \\
               SN 2009gs &  2009$-$06$-$15 &  $319.7060$ &   $-5.9530$ &                             ... &  $0.102$ &  Ia-norm &   3 &      ... &      ... &  ... \\
               SN 2009he &  2009$-$07$-$03 &  $245.5510$ &   $57.2729$ &                        $0.0306$ &  $0.008$ &  Ia-91bg &   1 &      ... &      ... &  ... \\
               SN 2009hi &  2009$-$07$-$10 &  $350.9840$ &   $16.7749$ &                        $0.0411$ &  $0.026$ &  Ia-norm &   3 &      ... &      ... &  ... \\
               SN 2009hk &  2009$-$07$-$11 &  $309.6560$ &  $-25.1156$ &  $0.0180$$^{\dagger ({\rm c})}$ &  $0.038$ &       Ia &   1 &      ... &      ... &  ... \\
               SN 2009hl &  2009$-$07$-$11 &  $262.7920$ &   $36.4278$ &  $0.0494$$^{\dagger ({\rm d})}$ &  $0.030$ &  Ia-norm &   2 &      ... &      ... &  ... \\
               SN 2009hn &  2009$-$07$-$24 &   $38.0013$ &    $1.2482$ &                        $0.0220$ &  $0.021$ &  Ia-norm &   1 &      ... &      ... &  ... \\
               SN 2009ho &  2009$-$07$-$25 &   $37.1389$ &   $37.9511$ &                             ... &  $0.049$ &       Ia &   1 &      ... &      ... &  ... \\
               SN 2009hp &  2009$-$07$-$26 &   $44.5998$ &    $6.5931$ &                        $0.0211$ &  $0.198$ &  Ia-norm &   1 &      ... &      ... &  ... \\
               SN 2009hs &  2009$-$07$-$28 &  $268.9620$ &   $62.5998$ &                        $0.0275$ &  $0.035$ &       Ia &   1 &    $8.6$ &      ... &    1 \\
               SN 2009hr &  2009$-$07$-$29 &   $10.1422$ &    $3.5414$ &  $0.0170$$^{\dagger ({\rm c})}$ &  $0.022$ &  Ia-norm &   1 &      ... &      ... &  ... \\
               PTF 09dlc &  2009$-$08$-$17 &  $326.6250$ &    $6.4192$ &  $0.0672$$^{\dagger ({\rm p})}$ &  $0.047$ &  Ia-norm &   2 &   $-2.3$ &   $18.2$ &    5 \\
               SN 2009jb &  2009$-$08$-$17 &  $260.9240$ &   $30.4971$ &  $0.0237$$^{\dagger ({\rm p})}$ &  $0.037$ &  Ia-norm &   3 &      ... &      ... &  ... \\
               PTF 09dnp &  2009$-$08$-$18 &  $229.8520$ &   $49.4990$ &  $0.0376$$^{\dagger ({\rm p})}$ &  $0.016$ &  Ia-norm &   2 &      ... &      ... &  ... \\
               SN 2009ig &  2009$-$08$-$20 &   $39.5484$ &   $-1.3125$ &                        $0.0088$ &  $0.028$ &  Ia-norm &  16 &  $-13.8$ &  $160.3$ &    1 \\
               SN 2009ih &  2009$-$08$-$21 &  $238.8790$ &   $41.9483$ &                        $0.0329$ &  $0.015$ &  Ia-91bg &   1 &      ... &      ... &  ... \\
               SN 2009ix &  2009$-$09$-$08 &   $49.4709$ &   $40.9589$ &                             ... &  $0.128$ &  Ia-norm &   2 &      ... &      ... &  ... \\
               SN 2009jg &  2009$-$09$-$22 &  $265.1430$ &   $18.7137$ &                             ... &  $0.062$ &  Ia-norm &   1 &      ... &      ... &  ... \\
               SN 2009jr &  2009$-$10$-$08 &  $306.6080$ &    $2.9092$ &                        $0.0165$ &  $0.116$ &  Ia-99aa &   2 &   $-3.6$ &    $5.3$ &    2 \\
               SN 2009jp &  2009$-$10$-$09 &  $349.4280$ &   $13.9569$ &  $0.0550$$^{\dagger ({\rm c})}$ &  $0.040$ &  Ia-norm &   1 &      ... &      ... &  ... \\
               SN 2009kk &  2009$-$10$-$15 &   $57.4345$ &   $-3.2644$ &                        $0.0129$ &  $0.118$ &  Ia-norm &   2 &   $-0.4$ &   $19.4$ &    2 \\
               SN 2009ko &  2009$-$10$-$28 &  $120.4930$ &   $15.0596$ &                        $0.0162$ &  $0.028$ &  Ia-norm &   2 &      ... &      ... &  ... \\
               SN 2009kq &  2009$-$11$-$05 &  $129.0630$ &   $28.0671$ &                        $0.0117$ &  $0.035$ &  Ia-norm &   4 &   $-9.2$ &   $28.4$ &    1 \\
               SN 2009lg &  2009$-$11$-$10 &  $354.7080$ &   $28.2651$ &                        $0.0580$ &  $0.165$ &  Ia-norm &   1 &      ... &      ... &  ... \\
               SN 2009le &  2009$-$11$-$16 &   $32.3214$ &  $-23.4124$ &                        $0.0178$ &  $0.014$ &  Ia-norm &   1 &   $17.5$ &      ... &    2 \\
               SN 2009li &  2009$-$11$-$16 &    $5.7142$ &    $6.9699$ &                        $0.0404$ &  $0.023$ &  Ia-norm &   1 &      ... &      ... &  ... \\
               SN 2009lv &  2009$-$11$-$19 &    $4.1107$ &   $22.4361$ &                             ... &  $0.059$ &  Ia-norm &   2 &      ... &      ... &  ... \\
               SN 2009lu &  2009$-$11$-$20 &  $163.5870$ &   $-4.3442$ &                        $0.0215$ &  $0.026$ &  Ia-norm &   1 &      ... &      ... &  ... \\
               SN 2009lr &  2009$-$11$-$23 &  $348.5600$ &   $-2.7533$ &                             ... &  $0.041$ &       Ia &   2 &      ... &      ... &  ... \\
               SN 2009me &  2009$-$12$-$03 &  $182.4160$ &   $43.6750$ &                             ... &  $0.012$ &  Ia-norm &   2 &      ... &      ... &  ... \\
               SN 2009mj &  2009$-$12$-$10 &  $103.3010$ &   $44.0713$ &                        $0.0196$ &  $0.092$ &      Iax &   1 &      ... &      ... &  ... \\
               SN 2009mh &  2009$-$12$-$12 &  $175.9830$ &   $10.7820$ &                        $0.0197$ &  $0.038$ &       Ia &   1 &      ... &      ... &  ... \\
               SN 2009mv &  2009$-$12$-$16 &  $108.9160$ &   $35.2412$ &                             ... &  $0.053$ &  Ia-norm &   1 &      ... &      ... &  ... \\
               SN 2009nr &  2009$-$12$-$22 &  $197.7460$ &   $11.4915$ &                        $0.0112$ &  $0.022$ &  Ia-norm &   3 &   $11.6$ &  $129.3$ &    6 \\
               SN 2009mz &  2009$-$12$-$26 &  $210.8530$ &   $-6.0587$ &                        $0.0086$ &  $0.024$ &  Ia-norm &   2 &      ... &      ... &  ... \\
               SN 2009na &  2009$-$12$-$26 &  $161.7560$ &   $26.5439$ &                        $0.0210$ &  $0.028$ &  Ia-norm &   3 &    $3.0$ &   $39.4$ &    2 \\
               SN 2009nq &  2009$-$12$-$28 &  $348.8210$ &   $19.0229$ &                        $0.0158$ &  $0.125$ &  Ia-norm &   1 &      ... &      ... &  ... \\
               SN 2009nk &  2009$-$12$-$29 &  $212.7450$ &    $6.3633$ &                        $0.0196$ &  $0.023$ &  Ia-norm &   2 &      ... &      ... &  ... \\
                SN 2010A &  2010$-$01$-$04 &   $38.1644$ &    $0.6195$ &                        $0.0207$ &  $0.025$ &  Ia-99aa &   1 &      ... &      ... &  ... \\
                SN 2010B &  2010$-$01$-$07 &  $208.5370$ &   $60.6804$ &                        $0.0102$ &  $0.011$ &  Ia-norm &   4 &      ... &      ... &  ... \\
                SN 2010N &  2010$-$01$-$12 &  $197.2720$ &   $17.0729$ &  $0.0210$$^{\dagger ({\rm c})}$ &  $0.019$ &  Ia-norm &   2 &      ... &      ... &  ... \\
                SN 2010H &  2010$-$01$-$16 &  $121.6020$ &    $1.0359$ &                        $0.0154$ &  $0.026$ &  Ia-norm &   3 &      ... &      ... &  ... \\
                SN 2010V &  2010$-$02$-$04 &  $217.1600$ &   $30.6360$ &  $0.0129$$^{\dagger ({\rm d})}$ &  $0.019$ &  Ia-norm &   3 &      ... &      ... &  ... \\
                SN 2010Y &  2010$-$02$-$08 &  $162.7660$ &   $65.7797$ &                        $0.0109$ &  $0.011$ &  Ia-norm &   4 &   $-6.4$ &   $22.3$ &    2 \\
               SN 2010p1 &  2010$-$02$-$12 &  $160.6750$ &   $58.8438$ &                        $0.0313$ &  $0.007$ &  Ia-norm &   2 &      ... &      ... &  ... \\
               SN 2010ag &  2010$-$03$-$05 &  $255.9730$ &   $31.5017$ &                        $0.0337$ &  $0.026$ &       Ia &   3 &    $0.3$ &   $58.3$ &    2 \\
               SN 2010ai &  2010$-$03$-$08 &  $194.8500$ &   $27.9964$ &  $0.0193$$^{\dagger ({\rm d})}$ &  $0.008$ &  Ia-norm &   1 &   $-6.4$ &      ... &    2 \\
               SN 2010an &  2010$-$03$-$11 &  $244.4190$ &   $35.0028$ &                        $0.0295$ &  $0.020$ &  Ia-norm &   3 &      ... &      ... &  ... \\
               SN 2010au &  2010$-$03$-$15 &  $138.1520$ &   $34.8547$ &                        $0.0615$ &  $0.018$ &  Ia-norm &   2 &      ... &      ... &  ... \\
               SN 2010ax &  2010$-$03$-$15 &  $220.4730$ &   $10.7504$ &                        $0.0508$ &  $0.024$ &  Ia-norm &   1 &      ... &      ... &  ... \\
               SN 2010ao &  2010$-$03$-$18 &  $205.9210$ &    $3.9000$ &                        $0.0228$ &  $0.023$ &  Ia-norm &   1 &  $-11.1$ &      ... &    1 \\
               SN 2010at &  2010$-$03$-$19 &  $181.2480$ &   $76.1312$ &                        $0.0418$ &  $0.073$ &       Ia &   1 &      ... &      ... &  ... \\
               SN 2010ba &  2010$-$03$-$21 &  $179.5860$ &   $15.3363$ &                             ... &  $0.037$ &  Ia-norm &   3 &      ... &      ... &  ... \\
               SN 2010bn &  2010$-$04$-$05 &  $176.2320$ &   $-5.0789$ &  $0.0530$$^{\dagger ({\rm c})}$ &  $0.019$ &  Ia-norm &   1 &      ... &      ... &  ... \\
               SN 2010bu &  2010$-$04$-$09 &  $235.7430$ &    $2.2813$ &  $0.0390$$^{\dagger ({\rm c})}$ &  $0.064$ &  Ia-norm &   2 &      ... &      ... &  ... \\
               SN 2010cp &  2010$-$05$-$09 &  $195.1080$ &  $-15.2889$ &                        $0.0164$ &  $0.049$ &  Ia-91bg &   1 &      ... &      ... &  ... \\
               SN 2010cs &  2010$-$05$-$12 &  $221.9820$ &   $19.0549$ &                        $0.0419$ &  $0.029$ &  Ia-norm &   1 &      ... &      ... &  ... \\
               SN 2010cr &  2010$-$05$-$15 &  $202.3540$ &   $11.7962$ &                        $0.0216$ &  $0.030$ &  Ia-norm &   2 &    $8.2$ &   $15.0$ &    2 \\
               SN 2010dl &  2010$-$05$-$24 &  $323.7540$ &   $-0.5133$ &                        $0.0300$ &  $0.033$ &  Ia-norm &   1 &   $18.1$ &      ... &    2 \\
               SN 2010eb &  2010$-$06$-$12 &   $20.4074$ &    $5.2944$ &                        $0.0076$ &  $0.026$ &  Ia-norm &   1 &      ... &      ... &  ... \\
               SN 2010gj &  2010$-$07$-$10 &  $327.7260$ &  $-17.7693$ &  $0.0370$$^{\dagger ({\rm c})}$ &  $0.047$ &       Ia &   1 &      ... &      ... &  ... \\
               SN 2010gl &  2010$-$07$-$18 &  $247.9110$ &   $59.6239$ &                        $0.0188$ &  $0.013$ &  Ia-norm &   2 &      ... &      ... &  ... \\
               SN 2010gv &  2010$-$08$-$09 &  $269.5940$ &   $50.7928$ &                             ... &  $0.040$ &       Ia &   1 &      ... &      ... &  ... \\
               SN 2010gz &  2010$-$08$-$16 &   $23.2128$ &  $-12.1893$ &                        $0.0184$ &  $0.021$ &  Ia-norm &   1 &      ... &      ... &  ... \\
               SN 2010hh &  2010$-$09$-$01 &  $269.8270$ &   $45.8756$ &                        $0.0190$ &  $0.033$ &  Ia-91bg &   1 &      ... &      ... &  ... \\
               SN 2010hz &  2010$-$09$-$12 &   $28.4249$ &   $29.9346$ &                        $0.0255$ &  $0.047$ &  Ia-norm &   1 &      ... &      ... &  ... \\
               SN 2010ii &  2010$-$09$-$30 &  $339.5550$ &   $35.4917$ &                        $0.0269$ &  $0.075$ &  Ia-norm &   2 &      ... &      ... &  ... \\
               SN 2010iw &  2010$-$10$-$14 &  $131.3130$ &   $27.8227$ &                        $0.0215$ &  $0.047$ &  Ia-norm &   1 &   $10.4$ &      ... &    2 \\
               SN 2010ju &  2010$-$11$-$14 &   $85.4833$ &   $18.4975$ &                        $0.0152$ &  $0.361$ &  Ia-norm &   2 &    $6.1$ &   $19.9$ &    1 \\
               SN 2010kg &  2010$-$11$-$29 &   $70.0350$ &    $7.3500$ &                        $0.0166$ &  $0.134$ &  Ia-norm &   2 &  $-13.0$ &    $0.8$ &    2 \\
                SN 2011H &  2011$-$01$-$04 &   $35.7751$ &   $43.0423$ &                        $0.0220$ &  $0.073$ &  Ia-norm &   1 &      ... &      ... &  ... \\
                SN 2011K &  2011$-$01$-$13 &   $71.3766$ &   $-7.3480$ &  $0.0145$$^{\dagger ({\rm d})}$ &  $0.088$ &  Ia-norm &   1 &    $9.1$ &      ... &    2 \\
                SN 2011U &  2011$-$01$-$28 &   $63.3914$ &   $27.5435$ &                        $0.0134$ &  $0.593$ &  Ia-norm &   1 &      ... &      ... &  ... \\
               SN 2011ao &  2011$-$03$-$03 &  $178.4630$ &   $33.3628$ &                        $0.0107$ &  $0.017$ &  Ia-norm &   3 &   $-8.7$ &   $37.6$ &    2 \\
               SN 2011ay &  2011$-$03$-$18 &  $105.6420$ &   $50.5903$ &                        $0.0210$ &  $0.072$ &      Iax &   9 &      ... &      ... &  ... \\
               SN 2011by &  2011$-$04$-$26 &  $178.9400$ &   $55.3261$ &                        $0.0028$ &  $0.012$ &  Ia-norm &  11 &  $-11.1$ &  $310.3$ &    1 \\
               SN 2011dm &  2011$-$06$-$15 &  $329.1730$ &   $73.2969$ &                        $0.0049$ &  $0.519$ &  Ia-norm &   1 &      ... &      ... &  ... \\
               SN 2011dn &  2011$-$06$-$21 &  $299.6480$ &    $2.6045$ &                        $0.0253$ &  $0.151$ &   Ia-pec &   1 &      ... &      ... &  ... \\
               SN 2011fg &  2011$-$08$-$20 &  $350.8360$ &   $16.7948$ &  $0.0450$$^{\dagger ({\rm d})}$ &  $0.023$ &  Ia-norm &   2 &      ... &      ... &  ... \\
               SN 2011fe &  2011$-$08$-$24 &  $210.7740$ &   $54.2737$ &                        $0.0008$ &  $0.008$ &  Ia-norm &  17 &  $-17.2$ &  $378.5$ &    1 \\
               SN 2011fk &  2011$-$08$-$29 &   $13.6753$ &   $36.7643$ &                        $0.0201$ &  $0.048$ &       Ia &   1 &      ... &      ... &  ... \\
               SN 2011fs &  2011$-$09$-$15 &  $334.3310$ &   $35.5806$ &                        $0.0209$ &  $0.101$ &  Ia-99aa &   3 &   $-2.6$ &   $25.8$ &    1 \\
               SN 2011gy &  2011$-$10$-$22 &   $52.3971$ &   $40.8676$ &  $0.0169$$^{\dagger ({\rm d})}$ &  $0.166$ &  Ia-norm &   1 &      ... &      ... &  ... \\
               SN 2011hb &  2011$-$10$-$24 &  $351.9810$ &    $8.7794$ &  $0.0289$$^{\dagger ({\rm d})}$ &  $0.051$ &  Ia-norm &   1 &      ... &      ... &  ... \\
               SN 2011iv &  2011$-$12$-$02 &   $54.7140$ &  $-35.5922$ &                        $0.0065$ &  $0.010$ &  Ia-norm &   2 &      ... &      ... &  ... \\
               SN 2011jh &  2011$-$12$-$22 &  $191.8100$ &  $-10.0631$ &                        $0.0078$ &  $0.032$ &  Ia-norm &   3 &      ... &      ... &  ... \\
               SN 2011jr &  2011$-$12$-$25 &  $106.6660$ &   $23.8936$ &                        $0.0226$ &  $0.052$ &  Ia-norm &   2 &      ... &      ... &  ... \\
               SN 2011jn &  2011$-$12$-$26 &  $194.3120$ &  $-17.4001$ &  $0.0475$$^{\dagger ({\rm d})}$ &  $0.059$ &  Ia-norm &   1 &      ... &      ... &  ... \\
               SN 2011jt &  2011$-$12$-$31 &  $223.3460$ &    $2.9620$ &  $0.0278$$^{\dagger ({\rm d})}$ &  $0.039$ &  Ia-norm &   2 &      ... &      ... &  ... \\
                SN 2012B &  2012$-$01$-$08 &   $57.8938$ &   $37.0785$ &  $0.0173$$^{\dagger ({\rm d})}$ &  $0.271$ &  Ia-norm &   1 &      ... &      ... &  ... \\
                SN 2012E &  2012$-$01$-$14 &   $38.3450$ &    $9.5849$ &                        $0.0203$ &  $0.063$ &  Ia-norm &   1 &   $-4.3$ &      ... &    1 \\
                SN 2012Z &  2012$-$01$-$29 &   $50.5223$ &  $-15.3877$ &                        $0.0071$ &  $0.034$ &      Iax &   4 &   $-7.6$ &   $35.1$ &    1 \\
               SN 2012c1 &  2012$-$03$-$27 &  $166.3340$ &   $-1.8681$ &                        $0.0908$ &  $0.047$ &   Ia-csm &   2 &      ... &      ... &  ... \\
               SN 2012cg &  2012$-$05$-$17 &  $186.8030$ &    $9.4203$ &                        $0.0015$ &  $0.018$ &  Ia-norm &  10 &  $-16.4$ &   $46.5$ &    1 \\
               SN 2012cu &  2012$-$06$-$14 &  $193.3720$ &    $2.1608$ &                        $0.0035$ &  $0.023$ &  Ia-norm &   3 &      ... &      ... &  ... \\
               SN 2012de &  2012$-$06$-$25 &  $333.7720$ &   $10.3035$ &                             ... &  $0.062$ &  Ia-norm &   1 &      ... &      ... &  ... \\
               SN 2012dn &  2012$-$07$-$08 &  $305.9010$ &  $-28.2787$ &  $0.0102$$^{\dagger ({\rm d})}$ &  $0.052$ &  Ia-norm &   2 &  $-14.6$ &   $-9.6$ &    1 \\
               SN 2012dv &  2012$-$07$-$18 &  $327.1260$ &  $-12.8392$ &                        $0.0700$ &  $0.037$ &  Ia-norm &   1 &      ... &      ... &  ... \\
               SN 2012ea &  2012$-$08$-$08 &  $266.2930$ &   $18.1408$ &                        $0.0102$ &  $0.055$ &       Ia &   2 &   $-6.8$ &   $29.8$ &    1 \\
               PTF 12ild &  2012$-$09$-$06 &  $338.2420$ &   $-0.2152$ &                        $0.1723$ &  $0.051$ &  Ia-norm &   1 &      ... &      ... &  ... \\
               PTF 12irf &  2012$-$09$-$15 &   $30.5316$ &    $0.1838$ &                        $0.1921$ &  $0.020$ &       Ia &   1 &      ... &      ... &  ... \\
               LSQ 12fhe &  2012$-$10$-$02 &  $323.0390$ &   $-5.7260$ &                        $0.0275$ &  $0.062$ &      Ia* &   1 &      ... &      ... &  ... \\
               SN 2012fr &  2012$-$10$-$27 &   $53.4000$ &  $-36.1271$ &                        $0.0055$ &  $0.018$ &  Ia-norm &   9 &   $-6.1$ &   $92.1$ &    7 \\
               SN 2012gl &  2012$-$10$-$29 &  $153.2100$ &   $12.6824$ &                        $0.0094$ &  $0.036$ &  Ia-norm &   1 &      ... &      ... &  ... \\
               SN 2012gx &  2012$-$11$-$18 &    $9.5073$ &  $-13.8610$ &  $0.0140$$^{\dagger ({\rm d})}$ &  $0.019$ &  Ia-norm &   1 &      ... &      ... &  ... \\
               SN 2012ht &  2012$-$12$-$18 &  $163.3450$ &   $16.7764$ &                        $0.0036$ &  $0.025$ &  Ia-norm &  13 &    $1.9$ &  $128.2$ &    8 \\
               SN 2012ij &  2012$-$12$-$29 &  $175.0660$ &   $17.4562$ &  $0.0110$$^{\dagger ({\rm d})}$ &  $0.023$ &  Ia-91bg &   4 &      ... &      ... &  ... \\
                SN 2013E &  2013$-$01$-$04 &  $150.0230$ &  $-34.2337$ &                        $0.0094$ &  $0.084$ &  Ia-norm &   5 &      ... &      ... &  ... \\
                SN 2013Q &  2013$-$01$-$25 &  $356.7830$ &   $29.4865$ &                        $0.0172$ &  $0.085$ &  Ia-norm &   2 &      ... &      ... &  ... \\
                SN 2013S &  2013$-$01$-$25 &   $53.8762$ &   $38.2832$ &  $0.0186$$^{\dagger ({\rm d})}$ &  $0.305$ &  Ia-99aa &   1 &      ... &      ... &  ... \\
               SN 2013gq &  2013$-$03$-$25 &  $124.4730$ &   $23.4696$ &                        $0.0139$ &  $0.049$ &  Ia-norm &   4 &    $1.1$ &   $16.7$ &    1 \\
               SN 2013ct &  2013$-$05$-$10 &   $18.2288$ &    $0.9794$ &                        $0.0038$ &  $0.024$ &       Ia &   1 &      ... &      ... &  ... \\
               SN 2013dj &  2013$-$06$-$10 &  $251.5080$ &    $6.4665$ &                        $0.0253$ &  $0.063$ &   Ia-91T &   1 &      ... &      ... &  ... \\
               SN 2013dh &  2013$-$06$-$12 &  $232.5050$ &   $12.9869$ &                        $0.0134$ &  $0.033$ &       Ia &   4 &   $-5.7$ &   $19.1$ &    1 \\
               SN 2013di &  2013$-$06$-$12 &  $339.1140$ &   $21.6151$ &                        $0.0238$ &  $0.040$ &  Ia-norm &   2 &      ... &      ... &  ... \\
               SN 2013dy &  2013$-$07$-$10 &  $334.5730$ &   $40.5693$ &                        $0.0039$ &  $0.132$ &  Ia-norm &  14 &  $-11.7$ &  $422.2$ &    1 \\
               SN 2013gh &  2013$-$08$-$08 &  $330.5910$ &  $-18.9168$ &                        $0.0088$ &  $0.025$ &      Ia* &   3 &  $-11.8$ &  $392.2$ &    1 \\
               SN 2013fa &  2013$-$08$-$25 &  $310.9730$ &   $12.5144$ &                        $0.0155$ &  $0.086$ &  Ia-norm &   1 &    $2.0$ &      ... &    1 \\
               SN 2013fw &  2013$-$10$-$21 &  $318.4370$ &   $13.5759$ &                        $0.0170$ &  $0.067$ &      Ia* &   1 &  $347.3$ &      ... &    1 \\
               SN 2013gs &  2013$-$11$-$29 &  $142.7870$ &   $46.3848$ &                        $0.0169$ &  $0.017$ &  Ia-norm &   1 &      ... &      ... &  ... \\
               SN 2013gy &  2013$-$12$-$06 &   $55.5703$ &   $-4.7218$ &                        $0.0140$ &  $0.050$ &  Ia-norm &   3 &    $5.6$ &  $271.8$ &    1 \\
   PSN J03055989+0432382 &  2013$-$12$-$21 &   $46.4995$ &    $4.5439$ &                             ... &  $0.146$ &  Ia-norm &   1 &      ... &      ... &  ... \\
               SN 2013hs &  2013$-$12$-$25 &   $29.7237$ &    $5.5904$ &                        $0.0194$ &  $0.036$ &  Ia-norm &   1 &      ... &      ... &  ... \\
                SN 2014J &  2014$-$01$-$21 &  $148.9260$ &   $69.6739$ &                        $0.0007$ &  $0.136$ &  Ia-norm &  10 &   $33.9$ &  $291.6$ &    1 \\
               SN 2014ag &  2014$-$03$-$11 &  $247.6690$ &   $44.5096$ &                        $0.0317$ &  $0.011$ &       Ia &   1 &      ... &      ... &  ... \\
               SN 2014ao &  2014$-$04$-$17 &  $128.6390$ &   $-2.5434$ &                        $0.0141$ &  $0.031$ &  Ia-norm &   1 &   $10.3$ &      ... &    1 \\
             ASASSN 14ar &  2014$-$04$-$24 &  $137.4240$ &   $37.6018$ &                        $0.0230$ &  $0.017$ &  Ia-norm &   1 &      ... &      ... &  ... \\
               SN 2014ck &  2014$-$06$-$29 &  $341.4120$ &   $73.1619$ &  $0.0050$$^{\dagger ({\rm d})}$ &  $0.394$ &      Iax &   2 &      ... &      ... &  ... \\
               SN 2014da &  2014$-$08$-$07 &    $7.3130$ &    $2.8660$ &  $0.0141$$^{\dagger ({\rm d})}$ &  $0.025$ &  Ia-91bg &   1 &      ... &      ... &  ... \\
             ASASSN 14gh &  2014$-$08$-$28 &  $258.7890$ &   $41.8109$ &  $0.0044$$^{\dagger ({\rm d})}$ &  $0.023$ &  Ia-norm &   1 &      ... &      ... &  ... \\
               SN 2014dg &  2014$-$09$-$11 &   $57.0824$ &   $70.1318$ &  $0.0040$$^{\dagger ({\rm d})}$ &  $0.628$ &  Ia-norm &  11 &      ... &      ... &  ... \\
               SN 2014dl &  2014$-$09$-$25 &  $247.4420$ &    $8.6418$ &                        $0.0330$ &  $0.054$ &   Ia-91T &   1 &      ... &      ... &  ... \\
               SN 2014dm &  2014$-$09$-$27 &   $62.0297$ &   $-8.8270$ &  $0.0330$$^{\dagger ({\rm d})}$ &  $0.041$ &  Ia-norm &   1 &      ... &      ... &  ... \\
               SN 2014dt &  2014$-$10$-$29 &  $185.4900$ &    $4.4718$ &                        $0.0052$ &  $0.019$ &      Iax &  13 &      ... &      ... &  ... \\
   PSN J03034759+0024146 &  2014$-$11$-$17 &   $45.9483$ &    $0.4041$ &                        $0.0430$ &  $0.073$ &  Ia-norm &   1 &      ... &      ... &  ... \\
              iPTF 14jfw &  2014$-$11$-$23 &  $137.5080$ &   $52.3157$ &                             ... &  $0.011$ &  Ia-norm &   1 &      ... &      ... &  ... \\
             ASASSN 14lp &  2014$-$12$-$09 &  $191.2880$ &    $0.4590$ &                        $0.0052$ &  $0.014$ &  Ia-norm &  15 &   $-0.8$ &  $175.1$ &    9 \\
              Gaia 15aba &  2015$-$02$-$06 &  $240.8760$ &   $52.2607$ &  $0.0460$$^{\dagger ({\rm d})}$ &  $0.015$ &  Ia-norm &   1 &      ... &      ... &  ... \\
              Gaia 15abu &  2015$-$02$-$09 &  $256.2090$ &   $41.0179$ &  $0.0750$$^{\dagger ({\rm d})}$ &  $0.024$ &  Ia-norm &   1 &      ... &      ... &  ... \\
              SNHunt 276 &  2015$-$02$-$10 &  $177.4950$ &   $21.3172$ &                        $0.0261$ &  $0.025$ &  Ia-91bg &   1 &      ... &      ... &  ... \\
                SN 2015H &  2015$-$02$-$10 &  $163.6760$ &  $-21.0705$ &                        $0.0125$ &  $0.047$ &      Iax &   1 &      ... &      ... &  ... \\
              Gaia 15aby &  2015$-$02$-$11 &  $214.8040$ &   $10.7169$ &  $0.0790$$^{\dagger ({\rm d})}$ &  $0.026$ &  Ia-norm &   1 &      ... &      ... &  ... \\
 PSN J13471211$-$2422171 &  2015$-$02$-$12 &  $206.8000$ &  $-24.3714$ &                        $0.0190$ &  $0.064$ &  Ia-norm &   1 &      ... &      ... &  ... \\
             ASASSN 15db &  2015$-$02$-$15 &  $236.7450$ &   $17.8840$ &                        $0.0113$ &  $0.029$ &  Ia-norm &   1 &      ... &      ... &  ... \\
              SNHunt 281 &  2015$-$03$-$16 &  $226.3670$ &    $1.6350$ &                        $0.0041$ &  $0.045$ &  Ia-norm &   3 &   $-5.3$ &   $20.5$ &   10 \\
             ASASSN 15fr &  2015$-$03$-$24 &  $140.0850$ &   $-7.6408$ &  $0.0334$$^{\dagger ({\rm d})}$ &  $0.033$ &  Ia-norm &   1 &      ... &      ... &  ... \\
             ASASSN 15hy &  2015$-$04$-$25 &  $302.5100$ &    $0.7392$ &  $0.0250$$^{\dagger ({\rm d})}$ &  $0.105$ &  Ia-norm &  12 &  $-13.4$ &  $152.2$ &    3 \\
             ASASSN 15jm &  2015$-$05$-$19 &  $260.2880$ &   $25.5821$ &                             ... &  $0.056$ &   Ia-csm &   1 &      ... &      ... &  ... \\
              iPTF 15awr &  2015$-$05$-$25 &  $225.3300$ &   $16.7800$ &                             ... &  $0.036$ &  Ia-norm &   1 &      ... &      ... &  ... \\
             ASASSN 15kx &  2015$-$06$-$10 &  $334.0490$ &   $37.4739$ &                        $0.0182$ &  $0.141$ &  Ia-norm &   3 &   $31.3$ &  $121.3$ &    3 \\
             ASASSN 15lo &  2015$-$06$-$19 &  $343.3910$ &   $19.7084$ &                             ... &  $0.056$ &  Ia-norm &   1 &      ... &      ... &  ... \\
             ASASSN 15lu &  2015$-$06$-$20 &  $200.3040$ &   $40.2658$ &                        $0.0350$ &  $0.014$ &  Ia-norm &   1 &   $-2.2$ &      ... &    3 \\
             ASASSN 15mc &  2015$-$07$-$05 &   $42.2482$ &    $3.1696$ &                        $0.0138$ &  $0.052$ &  Ia-norm &   3 &      ... &      ... &  ... \\
                SN 2015N &  2015$-$07$-$06 &  $325.8200$ &   $43.5799$ &                        $0.0149$ &  $0.456$ &  Ia-norm &  11 &   $-5.3$ &   $82.2$ &    1 \\
             ASASSN 15mi &  2015$-$07$-$06 &  $210.8160$ &   $41.6040$ &  $0.0344$$^{\dagger ({\rm d})}$ &  $0.018$ &  Ia-99aa &   1 &    $2.0$ &      ... &    3 \\
             ASASSN 15mg &  2015$-$07$-$09 &  $233.0950$ &   $41.8499$ &  $0.0428$$^{\dagger ({\rm d})}$ &  $0.028$ &  Ia-norm &   8 &   $-0.7$ &   $83.3$ &    3 \\
             ASASSN 15mp &  2015$-$07$-$17 &   $14.6886$ &  $-14.0699$ &  $0.0370$$^{\dagger ({\rm d})}$ &  $0.020$ &  Ia-99aa &   1 &      ... &      ... &  ... \\
               SN 2015ac &  2015$-$07$-$28 &  $349.1810$ &   $33.9966$ &                        $0.0168$ &  $0.062$ &       Ia &   1 &      ... &      ... &  ... \\
             ASASSN 15ns &  2015$-$08$-$06 &  $250.1170$ &   $39.3202$ &                        $0.0307$ &  $0.011$ &       Ia &   1 &      ... &      ... &  ... \\
             ASASSN 15og &  2015$-$08$-$13 &   $50.2810$ &  $-31.3127$ &  $0.0681$$^{\dagger ({\rm d})}$ &  $0.011$ &   Ia-csm &   7 &      ... &      ... &  ... \\
                PS 15cut &  2015$-$09$-$10 &  $358.1550$ &   $14.5526$ &                        $0.0266$ &  $0.036$ &  Ia-norm &   1 &      ... &      ... &  ... \\
   PSN J02524671+4656470 &  2015$-$09$-$12 &   $43.1946$ &   $46.9464$ &                        $0.0281$ &  $0.170$ &  Ia-norm &   1 &      ... &      ... &  ... \\
             ASASSN 15pr &  2015$-$09$-$13 &  $346.6650$ &  $-12.5729$ &  $0.0331$$^{\dagger ({\rm d})}$ &  $0.029$ &  Ia-norm &   1 &   $31.2$ &      ... &    3 \\
             ASASSN 15qc &  2015$-$10$-$01 &    $9.8249$ &    $3.9500$ &                        $0.0176$ &  $0.022$ &  Ia-norm &   1 &      ... &      ... &  ... \\
 MOT J041227.87+342902.0 &  2015$-$10$-$06 &   $63.1161$ &   $34.4839$ &                        $0.0214$ &  $0.312$ &  Ia-norm &   1 &      ... &      ... &  ... \\
                PS 15cku &  2015$-$10$-$16 &   $21.0939$ &    $3.5876$ &  $0.0230$$^{\dagger ({\rm d})}$ &  $0.023$ &  Ia-norm &   1 &   $-3.8$ &      ... &    3 \\
             ASASSN 15rm &  2015$-$10$-$19 &   $94.0160$ &  $-16.8249$ &                        $0.0208$ &  $0.147$ &  Ia-norm &   1 &      ... &      ... &  ... \\
             ASASSN 15rw &  2015$-$10$-$24 &   $33.9923$ &   $12.2374$ &  $0.0189$$^{\dagger ({\rm d})}$ &  $0.118$ &  Ia-norm &   1 &   $15.6$ &      ... &    3 \\
             ASASSN 15sf &  2015$-$10$-$30 &    $2.8650$ &   $-6.4273$ &  $0.0270$$^{\dagger ({\rm d})}$ &  $0.026$ &  Ia-norm &   2 &    $3.9$ &   $10.6$ &    3 \\
                 PS 16ud &  2015$-$11$-$01 &  $166.7820$ &   $-5.3789$ &                        $0.0373$ &  $0.063$ &  Ia-norm &   1 &      ... &      ... &  ... \\
             ASASSN 15so &  2015$-$11$-$08 &  $168.5460$ &   $48.3187$ &                        $0.0067$ &  $0.013$ &  Ia-norm &   2 &   $-7.7$ &   $27.0$ &    3 \\
   PSN J09100885+5003396 &  2015$-$11$-$08 &  $137.5370$ &   $50.0610$ &                        $0.0343$ &  $0.017$ &  Ia-norm &   7 &      ... &      ... &  ... \\
                PS 15cwx &  2015$-$11$-$17 &   $78.6992$ &    $7.0504$ &  $0.0460$$^{\dagger ({\rm d})}$ &  $0.149$ &  Ia-99aa &   1 &   $-3.0$ &      ... &    3 \\
               SN 2015bd &  2015$-$12$-$07 &  $170.9410$ &   $-1.1059$ &                        $0.0187$ &  $0.046$ &  Ia-norm &   5 &      ... &      ... &  ... \\
   PSN J12265018+1615496 &  2015$-$12$-$07 &  $186.7090$ &   $16.2638$ &  $0.0455$$^{\dagger ({\rm d})}$ &  $0.021$ &  Ia-norm &   1 &      ... &      ... &  ... \\
             ASASSN 15ub &  2015$-$12$-$14 &  $166.8040$ &   $65.0995$ &  $0.0320$$^{\dagger ({\rm d})}$ &  $0.014$ &  Ia-norm &   2 &      ... &      ... &  ... \\
                SN 2016F &  2016$-$01$-$04 &   $24.8835$ &   $33.8267$ &  $0.0161$$^{\dagger ({\rm d})}$ &  $0.042$ &       Ia &   1 &   $22.3$ &      ... &    3 \\
               SN 2016zc &  2016$-$01$-$28 &  $211.4880$ &   $43.8839$ &  $0.0337$$^{\dagger ({\rm d})}$ &  $0.006$ &  Ia-norm &   1 &      ... &      ... &  ... \\
              SN 2016aqt &  2016$-$02$-$28 &  $206.4610$ &   $26.7965$ &                             ... &  $0.015$ &  Ia-norm &   3 &      ... &      ... &    3 \\
              SN 2016blh &  2016$-$03$-$31 &  $212.4190$ &    $0.6567$ &  $0.0238$$^{\dagger ({\rm t})}$ &  $0.032$ &  Ia-norm &   1 &    $1.4$ &      ... &    3 \\
              SN 2016bln &  2016$-$04$-$04 &  $203.6900$ &   $13.8540$ &                        $0.0233$ &  $0.025$ &  Ia-norm &   3 &   $-4.0$ &   $24.2$ &    3 \\
              SN 2016bsa &  2016$-$04$-$22 &  $331.1480$ &   $42.3257$ &                        $0.0143$ &  $0.273$ &  Ia-norm &   1 &      ... &      ... &  ... \\
              SN 2016ccj &  2016$-$05$-$03 &  $257.6000$ &   $26.3966$ &  $0.0418$$^{\dagger ({\rm d})}$ &  $0.029$ &  Ia-norm &   1 &  $110.5$ &      ... &    3 \\
              SN 2016cmn &  2016$-$05$-$20 &  $277.5100$ &   $39.9655$ &                        $0.0183$ &  $0.053$ &  Ia-norm &   1 &      ... &      ... &  ... \\
              SN 2016coj &  2016$-$05$-$28 &  $182.0280$ &   $65.1773$ &                        $0.0045$ &  $0.016$ &  Ia-norm &  20 &  $-11.4$ &  $147.1$ &    1 \\
              SN 2016flv &  2016$-$08$-$27 &  $341.3750$ &   $-7.3353$ &  $0.0530$$^{\dagger ({\rm t})}$ &  $0.035$ &  Ia-norm &   2 &      ... &      ... &  ... \\
              SN 2016hvl &  2016$-$11$-$04 &  $101.0090$ &   $12.3966$ &  $0.0130$$^{\dagger ({\rm t})}$ &  $0.377$ &       Ia &   2 &   $16.6$ &  $104.5$ &    1 \\
              SN 2016ije &  2016$-$11$-$22 &   $29.6264$ &   $12.9244$ &                             ... &  $0.045$ &  Ia-91bg &   1 &      ... &      ... &  ... \\
              SN 2017cfd &  2017$-$03$-$16 &  $130.2050$ &   $73.4875$ &                        $0.0119$ &  $0.019$ &  Ia-norm &   4 &    $3.9$ &   $61.1$ &    1 \\
              SN 2017drh &  2017$-$05$-$03 &  $263.1090$ &    $7.0632$ &                        $0.0056$ &  $0.106$ &  Ia-norm &   7 &    $2.5$ &  $101.9$ &    1 \\
              SN 2017dws &  2017$-$05$-$03 &  $235.0590$ &   $11.3449$ &  $0.0818$$^{\dagger ({\rm t})}$ &  $0.035$ &  Ia-norm &   1 &    $7.7$ &      ... &    1 \\
              SN 2017dwp &  2017$-$05$-$04 &  $187.9320$ &   $36.2097$ &                        $0.0334$ &  $0.010$ &      Ia* &   3 &      ... &      ... &  ... \\
              SN 2017erp &  2017$-$06$-$13 &  $227.3120$ &  $-11.3342$ &                        $0.0063$ &  $0.093$ &  Ia-norm &  16 &   $-9.6$ &   $75.8$ &    1 \\
              SN 2017fgc &  2017$-$07$-$11 &   $20.0602$ &    $3.4028$ &                        $0.0081$ &  $0.029$ &  Ia-norm &  11 &   $-3.3$ &   $99.8$ &    1 \\
              SN 2017glx &  2017$-$09$-$03 &  $295.9180$ &   $56.1101$ &                        $0.0114$ &  $0.107$ &  Ia-norm &   4 &    $2.0$ &   $41.5$ &    1 \\
              SN 2017hbi &  2017$-$10$-$02 &   $38.1315$ &   $35.4836$ &                             ... &  $0.061$ &  Ia-norm &   5 &      ... &      ... &    1 \\
              SN 2017hou &  2017$-$10$-$24 &   $62.2589$ &   $-1.1601$ &                        $0.0167$ &  $0.108$ &  Ia-norm &   3 &      ... &      ... &  ... \\
              SN 2017hpa &  2017$-$10$-$25 &   $69.9615$ &    $7.0652$ &                        $0.0156$ &  $0.154$ &  Ia-norm &   5 &      ... &      ... &  ... \\
              SN 2017igr &  2017$-$11$-$18 &   $64.6815$ &   $26.9314$ &                        $0.0250$ &  $0.571$ &       Ia &   1 &      ... &      ... &  ... \\
              SN 2017iji &  2017$-$11$-$20 &  $183.1130$ &   $29.1493$ &                        $0.0135$ &  $0.018$ &  Ia-norm &   3 &      ... &      ... &  ... \\
              SN 2017iws &  2017$-$12$-$12 &  $130.6730$ &   $13.9678$ &  $0.0910$$^{\dagger ({\rm w})}$ &  $0.027$ &  Ia-norm &   1 &      ... &      ... &  ... \\
              SN 2017ixg &  2017$-$12$-$14 &  $350.1270$ &   $24.7776$ &  $0.0277$$^{\dagger ({\rm w})}$ &  $0.076$ &  Ia-norm &   2 &      ... &      ... &  ... \\
               SN 2018gl &  2018$-$01$-$13 &  $149.5260$ &   $10.3594$ &                        $0.0180$ &  $0.033$ &  Ia-norm &   1 &      ... &      ... &  ... \\
               SN 2018gv &  2018$-$01$-$15 &  $121.3940$ &  $-11.4379$ &                        $0.0053$ &  $0.050$ &  Ia-norm &   1 &    $8.7$ &      ... &    1 \\
               SN 2018pc &  2018$-$02$-$03 &  $142.2300$ &   $49.2381$ &                        $0.0090$ &  $0.012$ &  Ia-norm &   1 &      ... &      ... &  ... \\
               SN 2018pv &  2018$-$02$-$03 &  $178.2320$ &   $36.9866$ &                        $0.0031$ &  $0.018$ &       Ia &   1 &      ... &      ... &  ... \\
               SN 2018oh &  2018$-$02$-$04 &  $136.6650$ &   $19.3383$ &                        $0.0110$ &  $0.039$ &  Ia-norm &   1 &      ... &      ... &  ... \\
              SN 2018aae &  2018$-$02$-$06 &  $185.3920$ &   $55.5743$ &  $0.0290$$^{\dagger ({\rm d})}$ &  $0.010$ &  Ia-norm &   1 &      ... &      ... &  ... \\
              SN 2018aoz &  2018$-$04$-$02 &  $177.7580$ &  $-28.7441$ &                        $0.0058$ &  $0.072$ &  Ia-norm &   1 &   $37.7$ &      ... &    1 \\
              SN 2018bsn &  2018$-$05$-$14 &  $224.3700$ &    $5.8425$ &                        $0.0590$ &  $0.031$ &  Ia-norm &   1 &      ... &      ... &  ... \\
              SN 2018cni &  2018$-$06$-$13 &  $225.3450$ &  $-10.1805$ &  $0.0320$$^{\dagger ({\rm t})}$ &  $0.087$ &      Iax &   1 &      ... &      ... &  ... \\
              SN 2018eqq &  2018$-$08$-$03 &   $46.7298$ &   $41.5091$ &  $0.0160$$^{\dagger ({\rm t})}$ &  $0.127$ &       Ia &   1 &      ... &      ... &  ... \\
              SN 2018feb &  2018$-$08$-$16 &  $257.5470$ &   $21.6490$ &  $0.0148$$^{\dagger ({\rm t})}$ &  $0.052$ &  Ia-norm &   6 &      ... &      ... &  ... \\
              SN 2018hfp &  2018$-$10$-$07 &  $314.9490$ &  $-16.6369$ &                        $0.0291$ &  $0.063$ &  Ia-norm &   3 &      ... &      ... &  ... \\
              SN 2018hfr &  2018$-$10$-$10 &  $142.7300$ &   $-4.5712$ &                        $0.0226$ &  $0.024$ &   Ia-91T &   3 &      ... &      ... &  ... \\
              SN 2018hhn &  2018$-$10$-$13 &  $343.1340$ &   $11.6741$ &                        $0.0288$ &  $0.061$ &  Ia-norm &   3 &      ... &      ... &  ... \\
              SN 2018htt &  2018$-$10$-$31 &   $46.5121$ &  $-15.6116$ &  $0.0087$$^{\dagger ({\rm t})}$ &  $0.032$ &  Ia-norm &   1 &      ... &      ... &  ... \\
              SN 2018hzg &  2018$-$11$-$06 &  $175.5980$ &   $10.2644$ &                        $0.0216$ &  $0.049$ &       Ia &   1 &      ... &      ... &  ... \\
              SN 2018jaz &  2018$-$11$-$20 &  $204.8340$ &   $34.6888$ &  $0.0231$$^{\dagger ({\rm t})}$ &  $0.009$ &       Ia &   1 &      ... &      ... &  ... \\
\end{longtable}}
\twocolumn


\bsp	
\label{lastpage}
\end{document}